\documentclass[manuscript=article]{achemso}            		

\setkeys{acs}{usetitle=true}

\usepackage{graphicx}
\usepackage{subfigure}
\usepackage{multirow}
\usepackage[version=3]{mhchem}
\usepackage{longtable}
\usepackage{siunitx}
\usepackage{hyperref}
\usepackage{breakurl}
\usepackage{amsmath}
\DeclareSIUnit\angstrom{\text{\normalfont\AA}}
\usepackage{xr}

\title{Data-Efficient Construction of High-Fidelity Graph Deep Learning Interatomic Potentials}
\author{Tsz Wai Ko}
\email{t1ko@ucsd.edu}
\affiliation[UCSD]{Aiiso Yufeng Li Family Department of Chemical and Nano Engineering, University of California San Diego, 9500 Gilman Dr, Mail Code 0448, La Jolla, CA 92093-0448, United States}
\author{Shyue Ping Ong}
\email{ongsp@ucsd.edu} 
\affiliation[UCSD]{Aiiso Yufeng Li Family Department of Chemical and Nano Engineering, University of California San Diego, 9500 Gilman Dr, Mail Code 0448, La Jolla, CA 92093-0448, United States}
\date{}

\begin{document}
\maketitle

\begin{abstract}
Machine learning potentials (MLPs) have become an indispensable tool in large-scale atomistic simulations because of their ability to reproduce ab initio potential energy surfaces (PESs) very accurately at a fraction of computational cost. For computational efficiency, the training data for most MLPs today are computed using relatively cheap density functional theory (DFT) methods such as the Perdew-Burke-Ernzerhof (PBE) generalized gradient approximation (GGA) functional. Meta-GGAs such as the recently developed strongly constrained and appropriately normed (SCAN) functional have been shown to yield significantly improved descriptions of atomic interactions for diversely bonded systems, but their higher computational cost remains an impediment to their use in MLP development.
In this work, we outline a data-efficient multi-fidelity approach to constructing Materials 3-body Graph Network (M3GNet) interatomic potentials that integrate different levels of theory within a single model. Using silicon and water as examples, we show that a multi-fidelity M3GNet model trained on a combined dataset of low-fidelity GGA calculations with 10\% of high-fidelity SCAN calculations can achieve accuracies comparable to a single-fidelity M3GNet model trained on a dataset comprising 8$\times$ the number of SCAN calculations. This work paves the way for the development of high-fidelity MLPs in a cost-effective manner by leveraging existing low-fidelity datasets. 
\end{abstract}
		
\section{Introduction}
Atomistic simulations are an essential tool in materials science. A key input to atomistic simulations is an interatomic potential (IP) or force field that describes the potential energy surface (PES). Over the past decade, machine learning (ML) has emerged as a powerful approach to constructing IPs\cite{unke2021machine,  ko2021general, kocer2022neural} by learning the PES from a dataset of reference quantum mechanics (QM) calculations.\cite{behler2007generalized, ko2021fourth, ko2023accurate,bartok2010gaussian,thompson2015spectral,shapeev2016moment,drautz2019atomic} Such machine learning potentials (MLPs) typically achieve significantly improved accuracies in terms of energy and force errors over traditional empirical IPs\cite{zuo2020performance} and can be fitted more readily to more complex chemistries.\cite{li2020complex, lee2023atomic} MLPs have enabled studies inaccessible to either QM or empirical IPs, including the study of short-range chemical order and dislocation motion in high-entropy alloys\cite{kostiuchenko2019impact,santos2023short, byggmastar2021modeling}, long-time scale diffusion in lithium superionic conductors,\cite{qi2021bridging,krenzer2023nature,lacivita2018structural} etc. More recently, a new class of MLPs based on graph deep learning (GLPs) has been developed.\cite{chen2022universal,deng2023chgnet,batatia2022mace,batzner20223} GLPs utilize a natural graph representation for a collection of atoms, where the atoms are the nodes and the bonds between them are the edges. The information then passes through the graph via a series of message-passing steps, typically modeled using neural networks. The primary advantage of GLPs is their ability to handle arbitrary complex chemistries, without combinatorial explosion in the feature space associated with local environment-based MLPs\cite{ko2023recent}. Indeed, universal MLPs with coverage of the entire periodic table, such as the Materials 3-body Graph Network (M3GNet),\cite{chen2022universal} Crystal Hamiltonian Graph Network (CHGNet)\cite{deng2023chgnet} and Message-passing Atomic Cluster Expansion (MACE),\cite{batatia2022mace} have been trained using large public datasets of QM structural relaxations from the Materials Project\cite{jain2013commentary} and have broad applications in materials discovery and dynamic simulations.

The biggest bottleneck in MLP construction is the computation of the reference QM data set. For this reason, most MLPs today are fitted using data computed using relatively low-cost semi-local density functional theory (DFT) methods at the generalized gradient approximation (GGA) level, such as the Perdew-Burke-Ernzerhof (PBE) functional.\cite{perdew1996generalized} More accurate functionals exist. For instance, the strongly constrained and appropriately normed (SCAN) functional designed to recover all 17 exact constraints presently known for meta-GGA functionals has been demonstrated to yield significantly improved geometries and energies for diverse materials.\cite{sun2015strongly} However, such improved accuracies generally come at a higher computational expense.\cite{furness2020accurate} 
\begin{figure}
    \centering
    \includegraphics[width=1.0\textwidth]{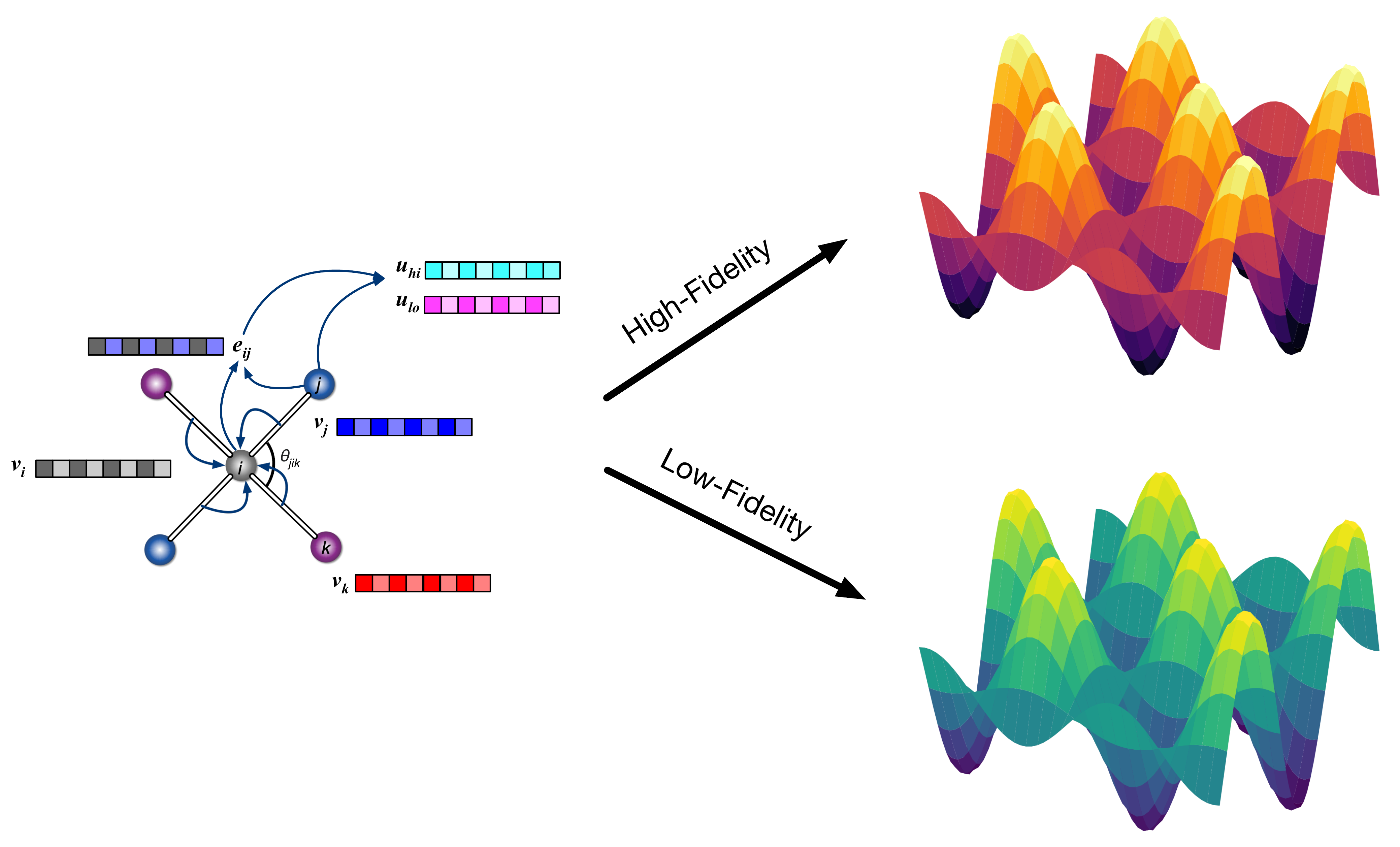}
    \caption{\textbf{Efficient construction of high-fidelity M3GNet interatomic potential.} Every structure is represented as a graph, considering atoms as nodes, bonds as edges, and the fidelity of reference methods as global state attributes. Atomic numbers and radial basis functions are embedded in latent node features and edge features, respectively. The fidelity of each data point is encoded as an integer (e.g. 0: low-fidelity, 1: high-fidelity in this work) to distinguish different levels of reference electronic structure methods. These integers are then embedded into the corresponding learnable state features. The information between nodes, edges and global state features are iteratively exchanged via sequential three-body interactions and graph convolutions. The resulting atomic features are fed into the readout layer yielding atomic energies. The summation of atomic contributions is equal to the potential energy of the system computed with the given fidelity method. }
    \label{fig:multi_m3gnet}
\end{figure}

Uniquely among the GLP architectures developed so far, the M3GNet architecture incorporates a global state feature (Figure \ref{fig:multi_m3gnet}). Previously, \citet{chen2021learning} have demonstrated the use of the global state feature for fidelity embedding to train a Materials Graph Network (MEGNet) model to predict band gaps using data from multiple fidelity sources (different functionals, experiments). The inclusion of a large quantity of low-fidelity (lofi) PBE band gaps was found to greatly enhance the resolution of latent structural features in MEGNet, which can significantly decrease the mean absolute errors (MAEs) of high-fidelity (hifi) band gap predictions by 22–45\%. Most critically, this improvement in accuracy was achieved without an increase in the amount of hifi training data.

In this work, we demonstrate the application of the multi-fidelity (mfi) approach to construct highly accurate M3GNet GLPs for two model systems - silicon and water. Both silicon and water have been extensively investigated with DFT at different levels of approximations\cite{cooper2000density,pedersen2017optimal, remsing2017dependence,remsing2018refined,cheng2019ab, ruiz2018quest,forster2014dispersion}, and the reported properties can be qualitatively different depending on the reference methods. For silicon, the radial distribution function (RDF) of liquid silicon computed with the SCAN functional is in better agreement with the experimental RDF due to its clearer discrimination of metallic and covalent bonds\cite{sun2016accurate, remsing2017dependence}. Furthermore, recent studies show that an MLP trained on SCAN data outperforms that trained on PBE data in terms of predictions of structural transitions from amorphous to polycrystalline phases at different external pressures\cite{deringer2021origins}, with a $\Delta-$learning MLP trained on more accurate random phase approximation (RPA) calculations as a reference\cite{bartok2018machine}. 
For water, meta-GGA and hybrid functionals such as SCAN~\cite{zheng2018structural, chen2017ab} and SCAN0\cite{zhang2021modeling} provide more accurate descriptions of hydrogen bonding and van der Waals interactions than GGA functionals such as revised-PBE (RPBE) and BLYP\cite{morawietz2016van, sprik1996ab, fernandez2005two}. We will demonstrate that with appropriate sampling, a mfi-M3GNet can be extremely efficient in terms of hifi data requirements, requiring 10\% coverage of SCAN data to achieve similar accuracies as an M3GNet GLP trained on an extensive SCAN dataset. We also provide comprehensive benchmarks reproducing the structural properties of liquid water and amorphous silicon, the Murnaghan equation of state, and the relaxed geometries of crystals. These results illustrate a data-efficient pathway to the construction of hifi GLPs. 

\section{Results}

\subsection{Multi-fidelity M3GNet architecture}

The M3GNet architecture has already been extensively covered in our previous work, and interested readers are referred to ref.\citenum{chen2022universal} for details. Here, we will discuss only the modifications for the treatment of data of multiple fidelities. For brevity, we will discuss the modifications in the context of datasets comprising only two fidelities, which is the most common scenario for PES datasets; extension beyond two fidelities is trivial.\cite{chen2021learning} The fidelity information (lofi RPBE/PBE and hifi SCAN) is encoded as integers (lofi: 0 and hifi: 1) and embedded as a vector in the global state feature input to the M3GNet model (Fig. \ref{fig:multi_m3gnet}).  This fidelity embedding encodes the complex functional relationship between different fidelities and their associated PESs and is automatically learned from the training dataset. The edges, nodes, and global state features are repeatedly updated through a series of interaction blocks that involve sequential three-body interactions and graph convolutions. The resulting atomic feature vectors, which represent the combined information of local chemical environments and fidelity, are fed into gated multi-layer perceptrons for the calculation of atomic energies.  

\begin{figure}
    \centering
    \includegraphics[width=1.0\textwidth]{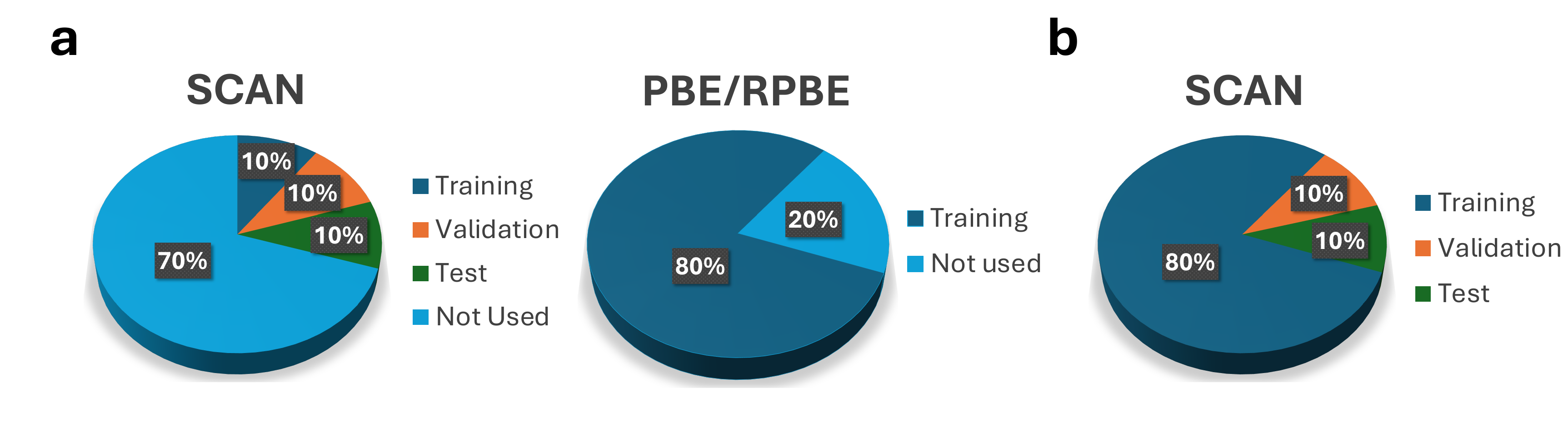}
    \caption{\textbf{Training data selection protocol for 10\%-mfi-SCAN and 80\%-1fi-SCAN models.} In the pie charts, similar positions correspond to similar structures. \textbf{a,} The training set (dark blue regions) for the 10\%-mfi SCAN model  comprises 10\% of the hifi SCAN data and 80\% of the lofi PBE/RPBE data. The validation (orange region) and test (green regions) sets are constructed by splitting equally the 20\% of the hifi SCAN data that do not appear in either the PBE/RPBE or SCAN training data. \textbf{b,} For the 80\%-1fi SCAN models, 80\% of SCAN data were used for training. The same structures were used for validation and test sets as the mfi SCAN model. }
    \label{fig:dataset_distribution}
\end{figure}

In the following sections, we will present the benchmarks for the performance of mfi M3GNet models for silicon and water against DFT and 1fi M3GNet models. Fig.~\ref{fig:dataset_distribution} illustrates the selection of lofi PBE/RPBE and hifi SCAN data for training, validation and test sets for constructing 10\%-mfi SCAN and 80\%-1fi SCAN models. As there is significant overlap in structures between the lofi and hifi SCAN datasets,\cite{deringer2021origins} only 80\% of the lofi data combined with 10\% of the hifi data \textit{selected from structures within the 80\% lofi data} were used to train the mfi models. The 20\% of SCAN data points, where the structures did not appear in both the lofi and hifi training sets, were divided into equal validation and test sets. This process avoids the possibility of including the same structures in the training and the validation/test sets, which allows for a robust comparison of different sampling sizes and techniques.

\subsection{Silicon}

In this section, we develop mfi and 1fi M3GNet IPs for silicon and benchmark their performance in reproducing not only basic energy and forces, but also structural properties of crystalline polymorphs and amorphous silicon as well as derived bulk properties such as the bulk modulus.

\subsubsection{Convergence with percentage of hifi data points}

\begin{figure}
    \centering
    \includegraphics[width=1.0\textwidth]{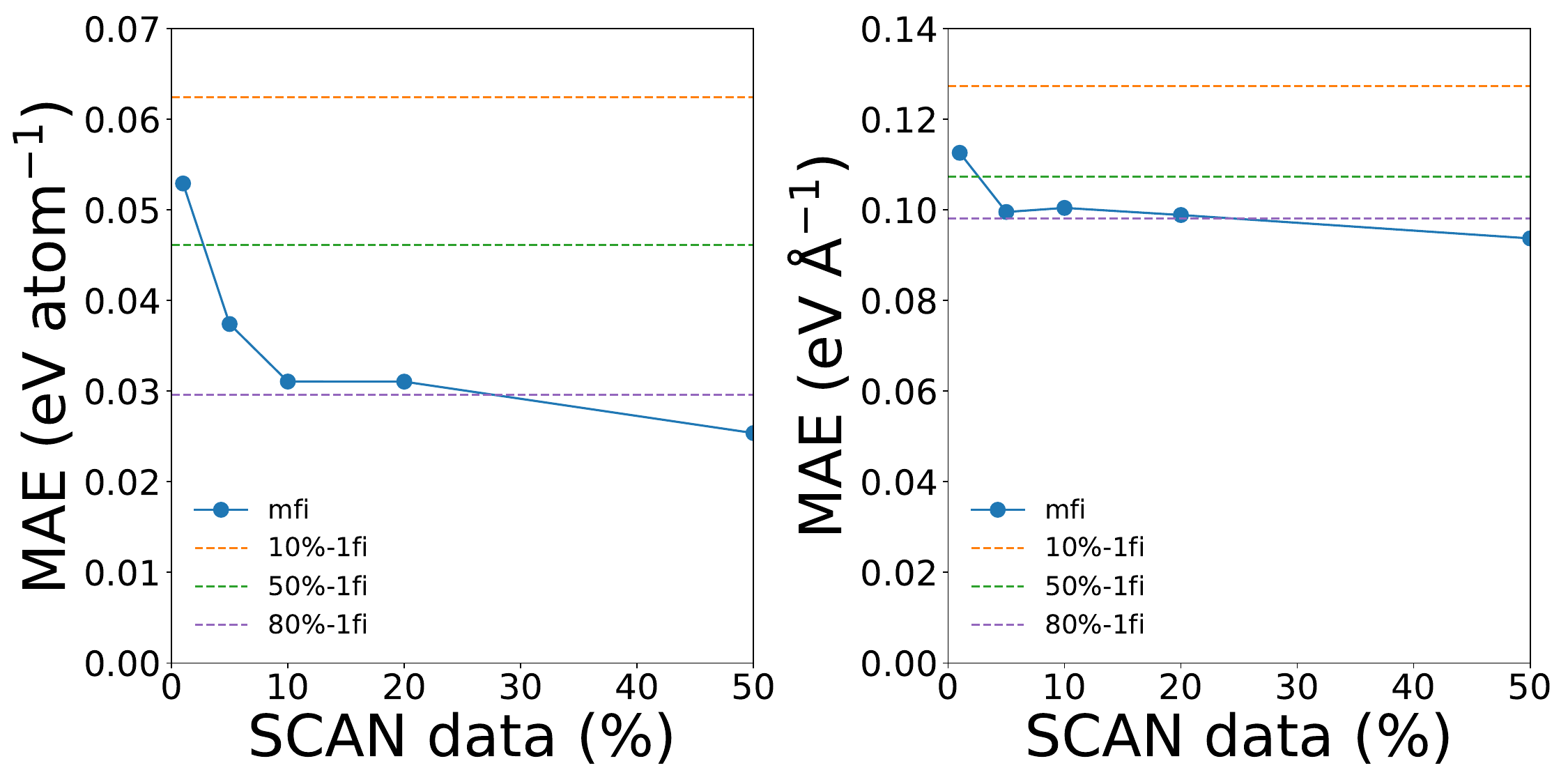}
    \caption{\textbf{Performance of multi-fidelity M3GNet potentials for silicon.} The test mean absolute errors (MAEs) of mfi M3GNet GLPs trained with different percentages of SCAN data for silicon. The dashed lines represent the MAEs of the 1fi M3GNet potential trained on 10\%, 50\% and 80\% of SCAN data for reference. }
    \label{fig:convergence_Si}
\end{figure}

Figure \ref{fig:convergence_Si} shows the convergence of the energy and force errors of the mfi M3GNet models for silicon with respect to the percentage of hifi (SCAN) data. Here, the Dimensionality-Reduced Encoded Cluster with Stratified (DIRECT) sampling approach developed by the current authors was used to ensure robust coverage of the configuration space regardless of the size of the SCAN dataset. With only 10\% of SCAN data, the mfi M3GNet model already achieves comparable energy and force MAEs as the 1fi M3GNet potential trained on the 80\% of SCAN data. Further decreases in MAEs with the increasing amount of SCAN training data are relatively small. Furthermore, we note that the MAEs of the 1fi M3GNet models are significantly higher than that of the mfi M3GNet models with comparable SCAN training data sizes. For example, the 10\%-1fi M3GNet model has energy and force MAEs of 0.062 eV and 0.127 eV \AA$^{-1}$, respectively, compared to the 10\%-mfi M3GNet model energy and force MAEs of 0.032 eV and 0.100 eV \AA$^{-1}$, respectively. Even for 50\% SCAN data, the mfi model significantly outperforms the 1fi model. Hence, the inclusion of the large quantity of lofi PBE data significantly improves the quality of hifi SCAN predictions. 

\subsubsection{Effect of sampling}

\begin{figure}
    \centering
    \includegraphics[width=0.5\textwidth]{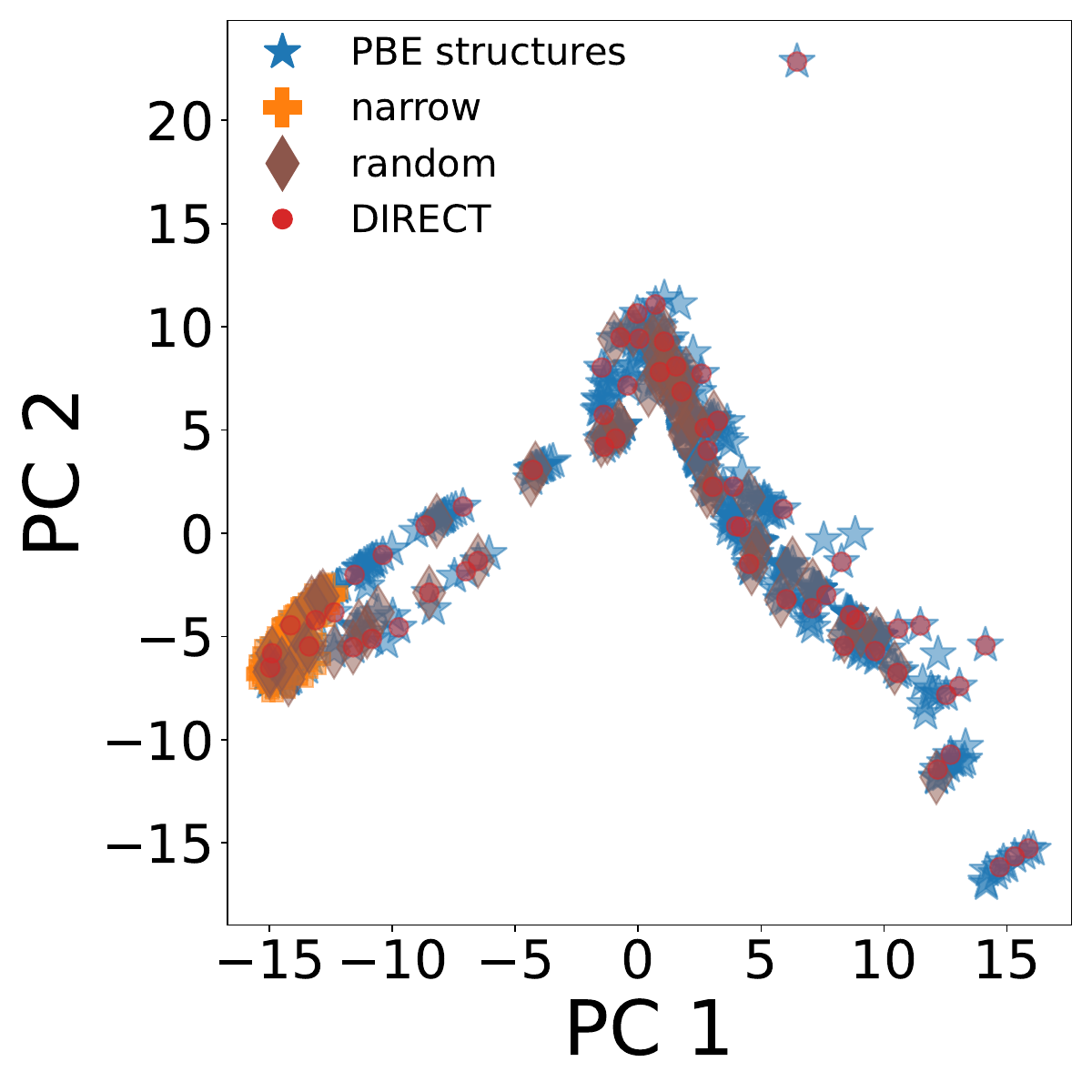}
    \caption{\textbf{Coverage of the first two principal components (PCs) of the M3GNet-encoded latent structure features for the three different sampling techniques.} The narrow sampling approach only samples structures from the far left of the PC space. The random sampling approach significantly improves the coverage of the entire space, but structures in the extrema of the latent space are missed. Finally, DIRECT sampling ensures coverage of the entire latent space, including extreme regions such as an isolated atom and highly distorted structures.}
    \label{fig:Coverage_Si}
\end{figure}

To explore the effect of sampling of hifi data on the performance of mfi M3GNet models, we trained 10\%-mfi M3GNet models, with the 10\% SCAN data sampled (a) from a narrow region (denoted as ``10\% mfi-narrow''), (b) in random manner (denoted as ``10\% mfi-random''), and (3) DIRECT sampling. The ``narrow'' sampling only includes primarily diamond crystal structures with distortions, strains and defects and excludes amorphous and other structures with more diverse local environments. Fig.~\ref{fig:Coverage_Si} illustrates the coverage of the three different sampling techniques using the first two principal components (PCs) of the encoded structure features, which were obtained using the pretrained M3GNet formation energy model as a structure featurizer. 

\begin{figure}
    \centering
    \includegraphics[width=1.0\textwidth]{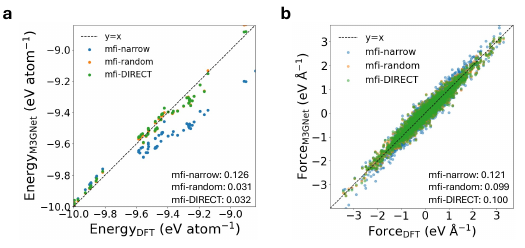}
    \caption{\textbf{Test mean absolute errors (MAEs) of energies and forces for M3GNet GLPs for silicon.} Parity plots of \textbf{a} energies and \textbf{b} forces predicted by 10\%-mfi-narrow, 10\%-mfi-random and 10\%-mfi-DIRECT. The numbers indicate the MAEs of energies in eV atom$^{-1}$ and forces in eV $\si{\angstrom}^{-1}$.}
    \label{fig:test_errors_Si}
\end{figure}

From Figure \ref{fig:test_errors_Si}, it can be observed that the 10\%-mfi-DIRECT and 10\%-mfi-random models exhibit similar test MAEs in energies and forces, but the 10\%-mfi-narrow model exhibits extremely high test MAEs.  These results illustrate that the 10\%-mfi-narrow model extrapolates poorly beyond the training domain for the majority of the validation and test structures.
This can be supported by the plot of the first two principal components of the training, validation, and test structures, as provided in Fig. S1. The validation errors, which share the same observations as the test errors, are plotted in Fig. S2.
To further evaluate the accuracy of 10\%-mfi-random and 10\%-mfi-DIRECT, we selected 297 bulk structures with great structural diversity from GAP-18 dataset\cite{bartok2018machine}, which is used for constructing general-purpose IPs of silicon, to compare the error of energies and forces with respect to DFT-SCAN. The MAEs in energies and forces of 10\%-mfi-random are 0.114 eV atom$^{-1}$ and 0.088 eV \si{\angstrom}$^{-1}$, respectively, while those of the 10\%-mfi-DIRECT are 0.040 eV atom$^{-1}$ and 0.091 eV \si{\angstrom}$^{-1}$, respectively. This clearly shows that the GLP achieves better accuracy with respect to unseen structures with DIRECT sampling.

\subsubsection{Equation of state of silicon polymorphs}

\begin{figure}
    \centering
    \includegraphics[width=1.0\textwidth]{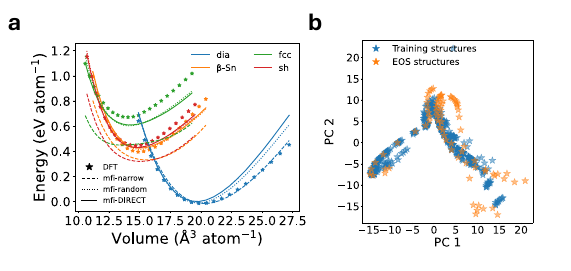}
    \caption{\textbf{Equation of state of silicon polymorphs from mfi M3GNet models.} \textbf{a,} Calculated DFT and mfi M3GNet energies of different silicon crystals as a function of the normalized volume. All energies are relative to the lowest DFT energy of the diamond structure. \textbf{b,} Visualization of the first two principal components of the encoded features using the M3GNet formation energy model for isotropically deformed crystals (``EOS structures'') and training structures. }
    \label{fig:eos_and_coverage}
\end{figure}
Figure \ref{fig:eos_and_coverage}a shows the calculated Murnaghan equation of state (EOS) for four silicon polymorphs. The EOS obtained from the 10\%-mfi-random and 10\%-mfi-DIRECT M3GNet models generally match DFT results in terms of the curvature and the energy minimum of crystals, even though the majority of the scaled crystals are outside the training domain (see Fig.~\ref{fig:eos_and_coverage}b). Table~\ref{tab:Si_EOS_prop} summarizes the equilibrium normalized volume, lowest energy and bulk modulus of crystals extracted from the fitted Murgnahan equation of state. 
Interestingly, the 10\%-mfi-narrow M3GNet has the smallest error in minimum energy and bulk modulus for the common diamond structure among all the models, but has much larger errors for the other silicon polymorphs. This is attributed to the comprehensive coverage of diamond structures and very limited coverage of other polymorphs by the 10\%-mfi-narrow GLP. In contrast, the 10\%-mfi-random and 10\%-mfi-DIRECT M3GNet GLPs provide considerably more diverse coverage in configuration space, leading to significantly improved prediction errors (especially for the bulk modulus) for the non-diamond structures, but with higher prediction errors for the diamond structure itself.

\begin{table}
    \centering
    \caption{\textbf{Equations of state for silicon polymorphs.} 
    The numbers reported are fitted values of the minimum energy ($E_0$), equilibrium volume ($V_0$) and bulk modulus ($B$) with Murnaghan equation of state. The numbers inside the parentheses indicate the mean absolute error for $E_0$ and absolute percentage error for $V_0$ and $B$ with respect to DFT.} 
    \begin{tabular}{c c c c}
    Crystals &  $E_0$ (eV atom$^{-1}$) & $V_0$ (\si{\angstrom}$^{3}$ atom$^{-1}$) & $B$ (GPa) \\
    \hline\hline
\textbf{Diamond} & & & \\
 mfi-narrow & -10.008 (0.002) &  20.0286 (1.45\%) &  102.092 (7.05\%) \\
 mfi-random & -10.000 (0.006) & 19.751 (1.23\%) & 118.101 (23.84\%) \\
 mfi-DIRECT & -9.899 (0.017) & 19.629 (1.84\%) & 125.178 (31.26\%)\\
    \hline
\textbf{\si{beta}-Sn} & & & \\
 mfi-narrow & -9.669 (0.074) &  15.403 (2.94\%) &  79.528 (31.06\%) \\
 mfi-random & -9.577 (0.018) &  15.052 (0.59\%) &  98.365 (14.73\%)\\
 mfi-DIRECT & -9.568 (0.027) &  15.073 (0.74\%) &  97.788 (15.23\%)\\
    \hline
\textbf{FCC} & & & \\
 mfi-narrow & -9.553(0.235) &   13.794 (1.28\%) &  48.805 (49.24\%) \\
 mfi-random & -9.390 (0.072) &  14.19  (1.58\%) &  91.839 (4.48\%) \\
 mfi-DIRECT & -9.386 (0.068) &  14.234 (1.87\%) &  82.718 (13.97\%)\\
    \hline
    \textbf{sh} & & & \\
 mfi-narrow & -9.679 (0.133) & 14.897 (1.22\%) & 72.821 (33.69\%)  \\
 mfi-random & -9.563 (0.016) & 14.823 (0.72\%) & 94.846 (13.64\%) \\
 mfi-DIRECT & -9.563 (0.017) & 14.853 (0.93\%) & 97.944 (10.82\%) \\
    \hline
    \end{tabular}
    \label{tab:Si_EOS_prop}
\end{table}

\subsubsection{Pressure-induced phase transition of amorphous silicon}

\begin{figure}
    \centering
    \includegraphics[width=1.0\textwidth]{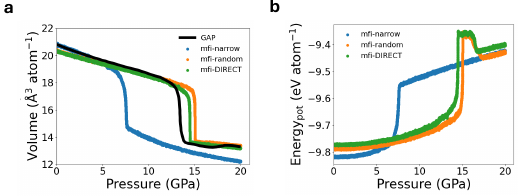}
    \caption{\textbf{Structural transition of amorphous silicon induced by external pressurization.} The evolution of \textbf{a} normalized volume and \textbf{b} potential energy of amorphous silicon with 1728 atoms is simulated by increasing the pressurization rate = 0.5 GPa ps$^{-1}$ from 0 GPa to 20 GPa. The simulated volume using Gaussian approximation potential (GAP) is reproduced from ref.\citenum{deringer2021origins}. }
    \label{fig:aSi_phase_trans}
\end{figure}

Amorphous silicon is known to undergo pressure-induced phase transitions. Here, isothermal compression MD simulations were performed on amorphous silicon cells containing 1728 atoms (see Methods section). Fig.\ref{fig:aSi_phase_trans}a shows the volume change of amorphous silicon under a uniform pressurization rate of 0.5 GPa ps$^{-1}$ with the various M3GNet GLPs. It can be observed that MD simulations using both 10\%-mfi-random and 10\%-mfi-DIRECT successfully reproduce the phase transitions observed in previous work using MD simulations with a Gaussian Approximation Potential (GAP) on 100,000 amorphous silicon atoms.\cite{deringer2021origins} For instance, a volume collapse to very-high-density amorphous (VHDA) states is observed at $\sim 15$ GPa, and recrystalliization is observed with further compression. MD simulations with the 10\%-mfi-narrow M3GNet GLP, on the other hand, predicts the structural collapse to occur at a much lower pressure ($\sim 7$ GPa), and no recrystallization was observed with further compression. From Fig.\ref{fig:aSi_phase_trans}b, we can observe a sharp increase in potential energy at the phase transition pressure. This is followed by a small drop in potential energy with recrystallization in the 10\%-mfi-random and 10\%-mfi-DIRECT simulations, while no such drop was observed in the 10\%-mfi-narrow simulations. 

\begin{figure}
    \centering
    \includegraphics[width=1.0\textwidth]{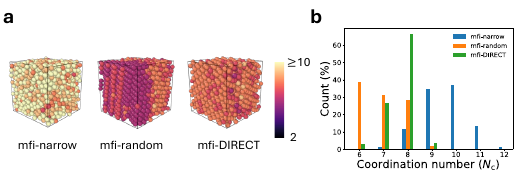}
    \caption{\textbf{Coordination analysis of disordered silicon at 20 GPa and 500 K.} \textbf{a,} The snapshot of isothermal compressed silicon structures. The heat map indicates the coordination number with a spatial cutoff = 2.85 \si{\angstrom}. \textbf{b,} The statistic of coordination number is obtained from 100 ps \textit{NPT} simulations. The structures with color coding were visualized using Ovito\cite{stukowski2009visualization}.}
    \label{fig:aSi_coord_barplot}
\end{figure}

Fig.\ref{fig:aSi_coord_barplot} shows that the local environments of structures extracted from simulations under 20 GPa at 500 K using all three GLPs. From Fig.\ref{fig:aSi_coord_barplot}b, it can be clearly observed that the structures extracted from the 10\%-mfi-random and 10\%-mfi-DIRECT simulations exhibit a clearly crystalline nature with lower coordination numbers ($7 \leq N_c \leq 8$), while the structure extracted from the 10\%-mfi-narrow simulation appears to remain amorphous with a greater percentage of highly coordinated Si atoms ($N_c > 8$). This poor performance of the 10\%-mfi-narrow GLP can be attributed to the underestimation of short-range repulsive forces as can be seen in the relatively flat curvature of EOS curves (Fig.\ref{fig:eos_and_coverage}b) and the underestimation of the energy of high energy structures (Fig.\ref{fig:test_errors_Si}a).

Interestingly, the 10\%-mfi-random structure contains about 40\% of atoms in a 6-fold-coordinated $\beta$-Sn-like phase (and 30\% as 7-fold coordinated and 8-fold-coordinated simple hexagonal (sh)-like phase.  The 10\%-mfi-DIRECT structure contains around 65\% of atoms as 8-fold-coordinated sh crystallites, 25\% in 7-fold coordination, and the remaining atoms in a 6-fold-coordinated $\beta$-Sn-like phase. Some evidence of phase segregation is also observed in the 10\%-mfi-random structure while no such phase segregation is observed in the 10\%-mfi-DIRECT structure. We note that the 10\%-mfi-DIRECT GLP is expected to better sample the extrema of the configuration space compared to the 10\%-mfi-random GLP. As a result, the transition pressure predicted using the 10\%-mfi-DIRECT is somewhat closer to the previous GAP results (Fig.\ref{fig:aSi_phase_trans}a), and the eventual recrystallized structure corresponds better to the crystalline counterpart of the VHDA form (sh crystal).

\subsection{Water}

In this section, we develop mfi and 1fi M3GNet GLPs for water. Given the relatively similar performances of DIRECT and random sampling for the mfi M3GNet GLPs for silicon, only random sampling was used to select hifi training data for water.

\subsubsection{Energy and force errors}

\begin{figure}
    \centering
    \includegraphics[width=1.0\textwidth]{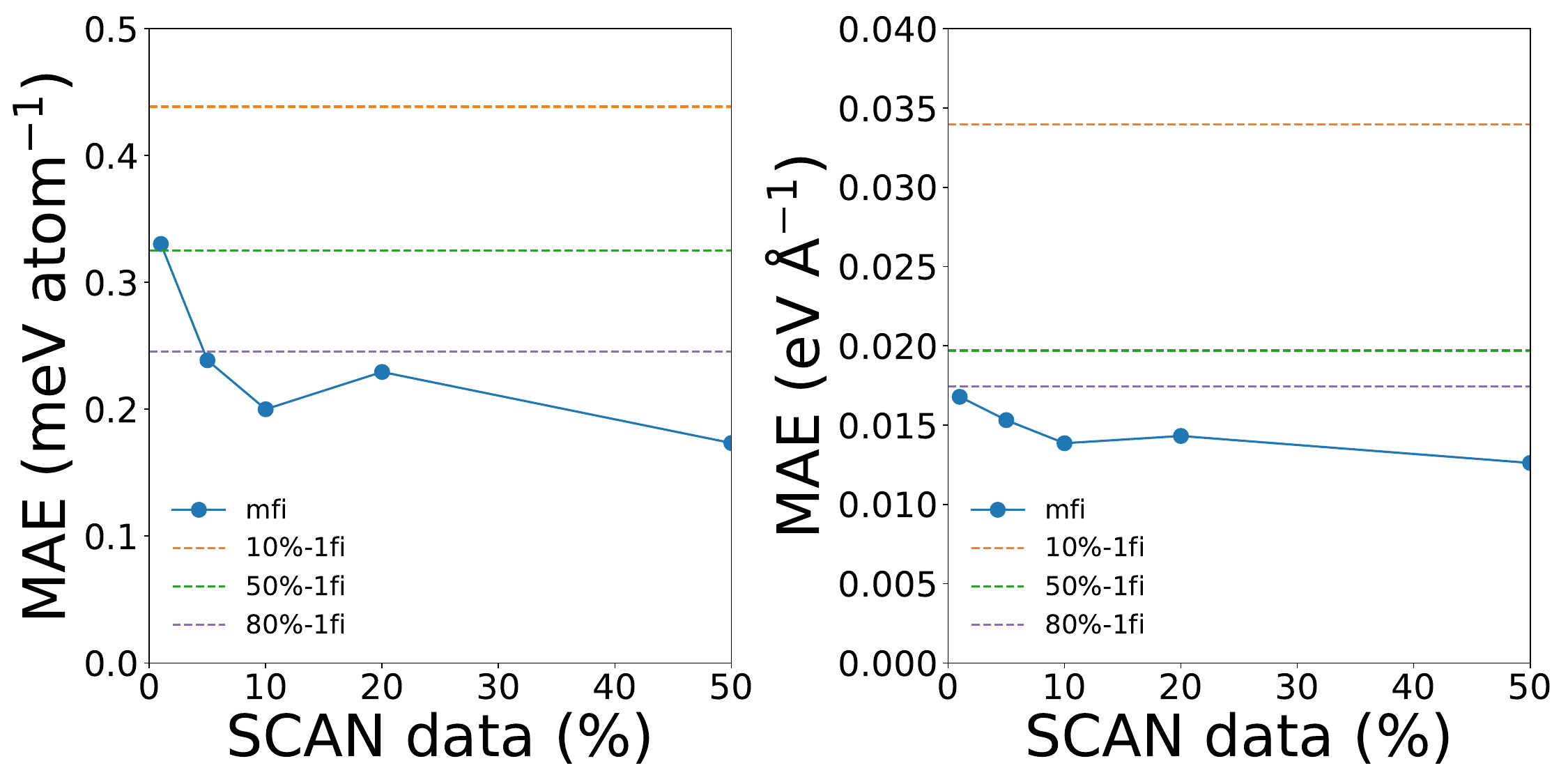}
    \caption{\textbf{Performance of mfi M3GNet potentials for water.} The test mean absolute errors (MAEs) of mfi M3GNet GLPs trained with different percentages of SCAN data for silicon. The dashed lines represent the MAEs of the 1fi M3GNet potential trained on 10\%, 50\% and 80\% of SCAN data for reference. }
    \label{fig:convergence_H2O}
\end{figure}

Fig.\ref{fig:convergence_H2O} shows the convergence of the energy and force MAEs of mfi M3GNet models for water with respect to the percentage of hifi (SCAN) data. The parity plots of energies and forces for all GLPs are provided in Fig. S4 and S5. The 10\%-mfi M3GNet has energy and force MAEs of 0.2 meV atom$^{-1}$ and 0.014 eV $\si{\angstrom}^{-1}$, respectively, which is less than half those of the 10\%-1fi model and outperforms even the 80\%-1fi M3GNet GLP.

\subsubsection{Structure of liquid water}
\begin{figure}
    \centering
    \includegraphics[width=\textwidth]{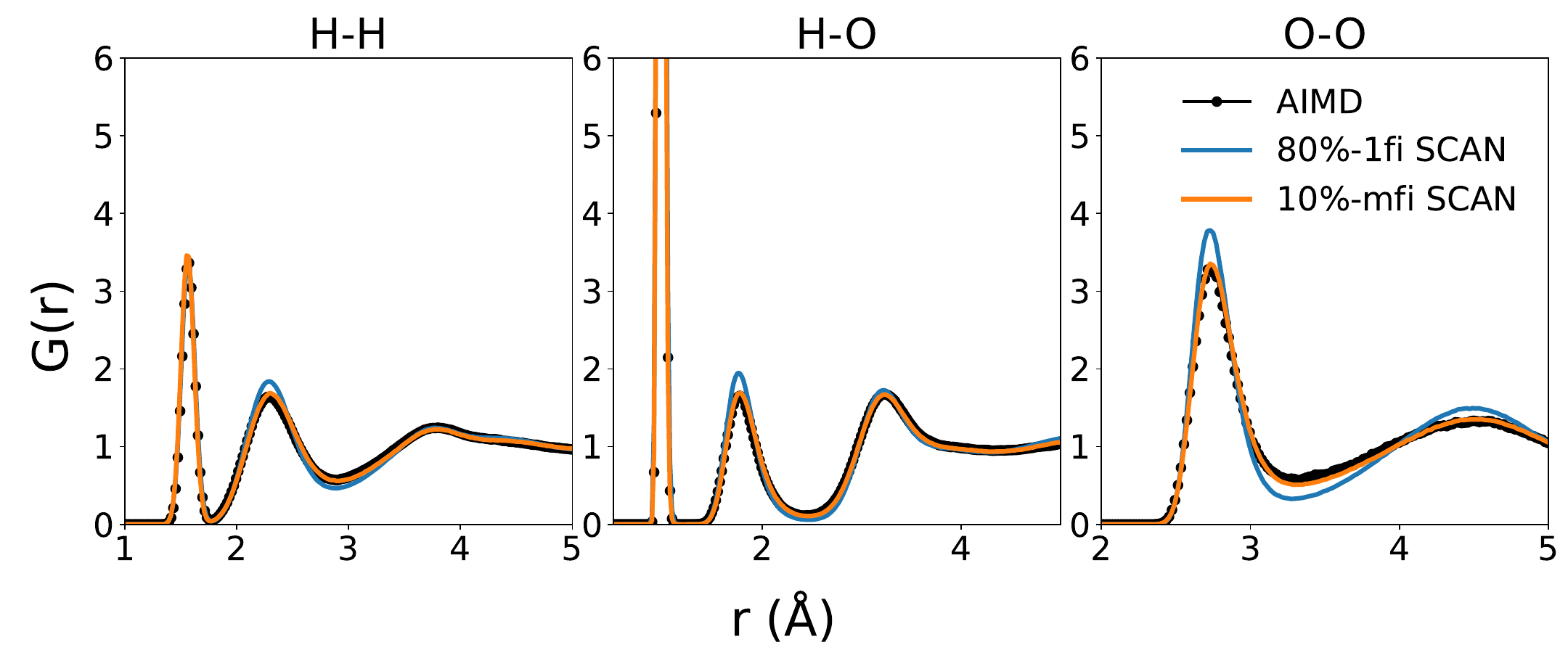}
    \caption{\textbf{Radial distribution function (RDF) of liquid water.} The RDFs were extracted from NPT simulations with the 80\%-1fi and 10\%-mfi M3GNet GLPs at 300K and 1 atm. The SCAN-AIMD results were obtained from ref.\citenum{yao2020temperature}.}
    \label{fig:H2O_RDF}
\end{figure}
\textit{NPT} MD simulations were performed on a model containing 64 formula units of \ce{H2O} using the 10\%-mfi and 1fi M3GNet IPs. Fig.~\ref{fig:H2O_RDF} compares the radial distribution functions (RDFs) of liquid water at 300K obtained using the mfi M3GNet GLPs and prior SCAN AIMD results. The 10\%-mfi SCAN model is in agreement with 80\%-1fi SCAN in terms of the position and magnitude of peaks. Notably, 10\%-mfi SCAN predicted a lower first peak in the RDF of oxygen-oxygen, which better aligns with SCAN-AIMD results\cite{yao2020temperature} compared to 80\%-1fi SCAN. Similarly, the 10\%-mfi SCAN generally matches the shape of the angular distribution function (ADF) of O-O-O with 80\%-1fi SCAN. Notably, the magnitude of peaks in the ADF predicted by 10\%-mfi SCAN is even in better agreement with SCAN-AIMD\cite{yao2020temperature}. All these analyses show that the mfi M3GNet model enables the efficient construction of hifi MLPs with state-of-the-art accuracy using a fraction of data points, where the lofi dataset provides additional coverage in configuration space as shown in the first two PCs of reduced structural latent features from Fig. S6. The RDFs and ADFs at other temperatures can be also found in Fig. S7 and Fig. S8, respectively.
\begin{figure}[H]
    \centering
    \includegraphics[width=0.5\textwidth]{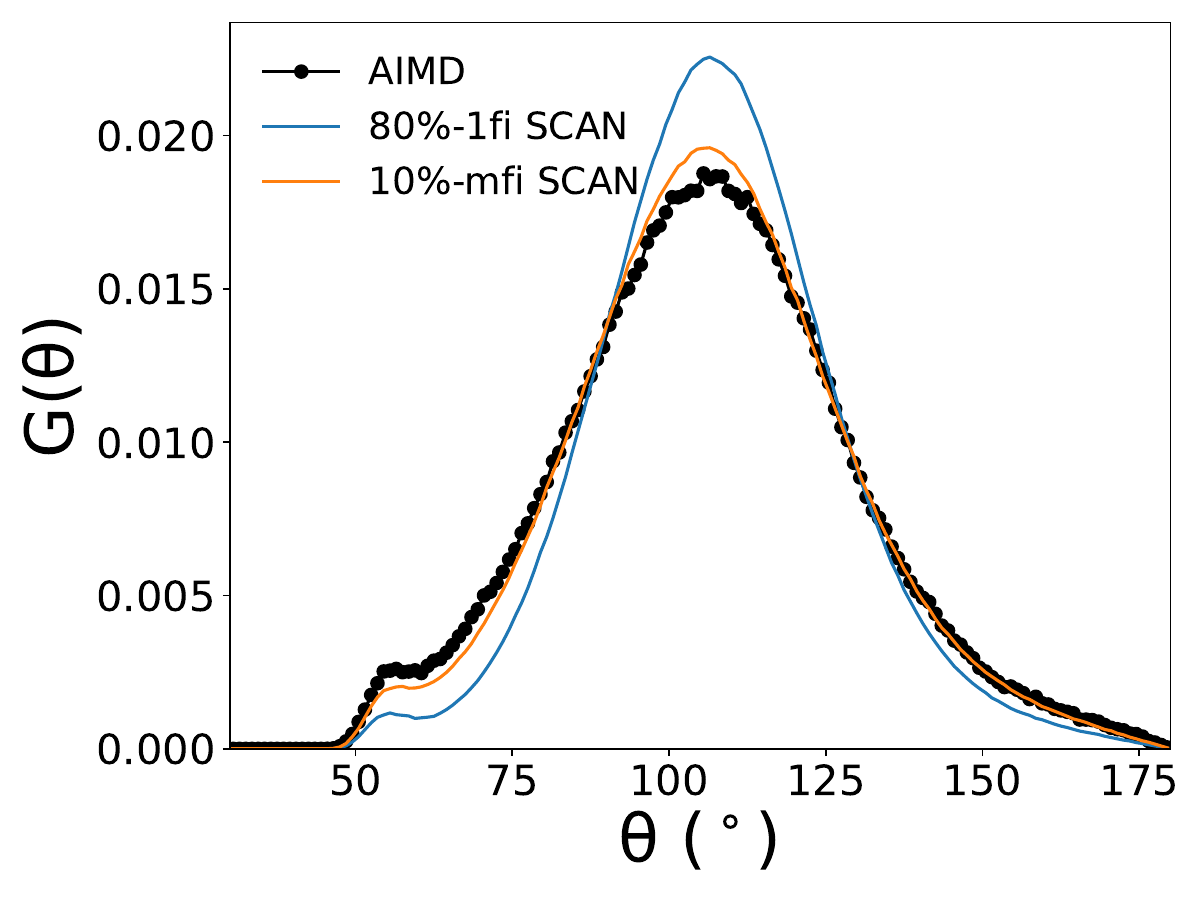}
    \caption{\textbf{Angular Distribution Function of Liquid Water.} The O-O-O triplet angular distribution function of liquid water was computed with 80\%-1fi SCAN and 10\%-mfi SCAN at 300K and 1 atm. The AIMD results were obtained from ref.\citenum{yao2020temperature}.}
    \label{fig:H2O_ADF}
\end{figure}

\subsubsection{Structure of ice polymorphs}
\begin{figure}
    \centering
    \includegraphics[width=0.5\textwidth]{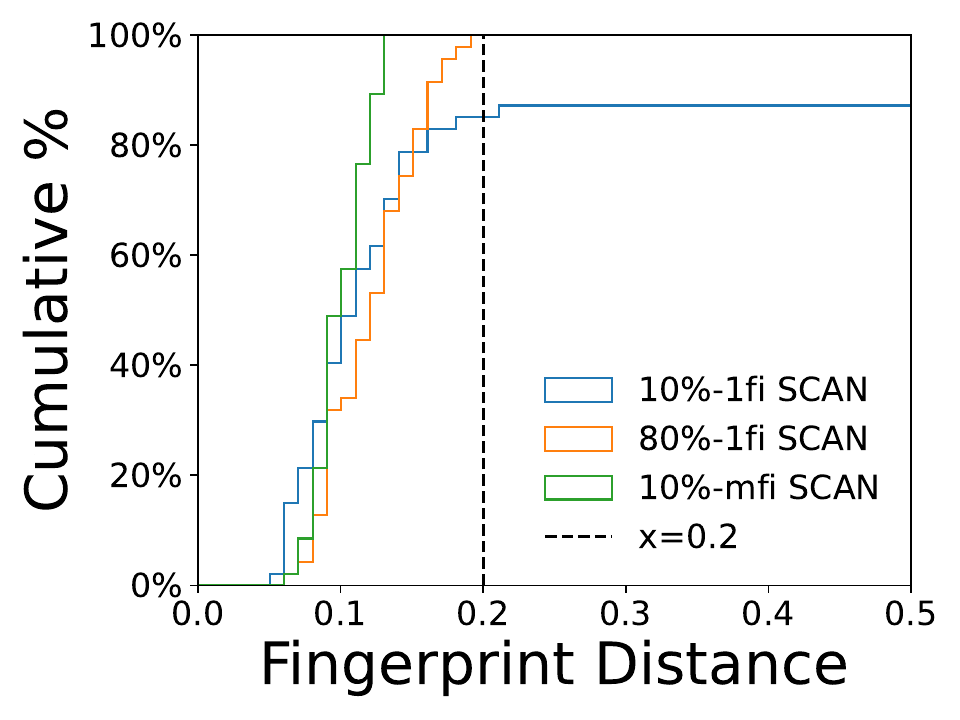}
    \caption{\textbf{Distribution of structural fingerprint distances for ice crystals relaxed with M3GNet and SCAN.} Cumulative histogram of calculated CrystalNN fingerprint distance between different M3GNet- and SCAN-relaxed ice structures. All these crystals are obtained from ref.\citenum{monserrat2020liquid}. The legend indicates the fraction of data points used for training. A smaller fingerprint distance means greater similarity between two relaxed structures and the vertical dashed line is used to visualize the fraction of structures that fall below 0.2 in terms of fingerprint distance. }
    \label{fig:RMSD_H2O}
\end{figure} 
\par We also compared the performance of 10\%-1fi-SCAN, 80\%-1fi-SCAN and 10\%-mfi-SCAN M3GNet with DFT-SCAN by conducting geometry relaxation on various small ice crystals ($N_{atoms}<100$) reported in ref.\citenum{monserrat2020liquid}, which covers the majority of experimentally known phases. 
To measure the similarity between two crystal structures, we employed CrystalNN algorithm\cite{pan2021benchmarking} to compute the structural fingerprint of periodic structures based on the Voronoi algorithm combined with the solid angle weights to determine the probability of various coordination environments.
The smaller fingerprint distance indicates the higher similarity between the two crystal structures. Fig.~\ref{fig:RMSD_H2O} shows the distribution of fingerprint distance among three different M3GNet models. The 10\%-mfi SCAN generally achieves the lowest fingerprint distance between DFT-relaxed and M3GNet-relaxed structures among the three models. The main reason is that the lofi training set covers several crystal structures, which allows the mfi model to provide a more informative representation of local environments for other crystals resulting in better agreement with DFT-SCAN. In contrast, the other two models only contain the liquid water structures for training and therefore the extrapolation limits the accuracy of both models.
The main reason is that the lofi training set covers various crystal structures. This enables the mfi model to offer a more informative representation of local environments for other crystals, resulting in better agreement with DFT-SCAN.
In contrast, the training set for 1fi models only includes liquid water structures for training, which provides less accurate predictions for crystals.

\section{Discussion}

In the preceding sections, we have demonstrated that a multi-fidelity graph network architecture provides a data-efficient pathway to construct GLPs for higher levels of theory. A GLP trained on a dataset comprising a large quantity of lofi PBE/RPBE data with a fraction of hifi SCAN data can achieve accuracies comparable to that of a GLP trained on a much larger hifi SCAN dataset. 

\begin{figure}
    \centering
    \includegraphics[width=1.0\textwidth]{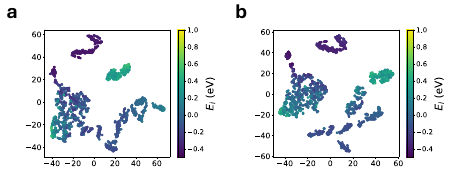}
    \caption{\textbf{Two-dimensional t-distributed Stochastic Neighbor Embedding of Atomic Features for Silicon.} The atomic features for the test set of silicon were extracted from \textbf{a} 10\%-1fi-DIRECT, and \textbf{b} 10\%-mfi-DIRECT. }
    \label{fig:Si_TSNE}
\end{figure}
Fig.~\ref{fig:Si_TSNE} compares the t-distributed stochastic neighbor embedding (T-SNE) analysis of the atomic features extracted from the silicon test set from a mfi and 1fi M3GNet trained on a 10\% DIRECT SCAN data. Consistent with prior analyses, the mfi M3GNet model exhibits better separation between structures of high and low atomic energies in the latent space compared to the 1fi M3GNet model. We surmise that it is this improvement in latent representation with the inclusion of a large lofi PBE dataset that led to the improvement in the performance of the mfi M3GNet model.

Though we have demonstrated the multi-fidelity concept using two relatively simple model systems (silicon and water), the impact of this work goes beyond custom MLPs for specific chemistries. For instance, a major area of active research is in the development of universal GLPs (also so-called ``foundational'' MLPs) with broad coverage of the periodic table of the elements. Generating a robust large dataset, even with standard GGA methods, for such universal GLPs is a major challenge. Using the multi-fidelity approach outlined in this work, we anticipate that training universal GLPs based on state-of-the-art meta-GGAs such as the SCAN functional can be made significantly more data-efficient. This approach can also be extended to other GLP architectures through appropriate modification to include a global fidelity embedding feature and additional message-passing operations.

\section{Methods}
\subsection{M3GNet architecture}

The Materials 3-body Graph Network (M3GNet) model architecture has been extensively covered previously, and interested readers are referred to ref.\citenum{chen2022universal} for details. Only the major parameter settings used in this work will be discussed here. The distance cut-off that defines the edges $\mathbf{e}_{ij}$ and the angles between two edges $\mathbf{e}_{ij}$ and $\mathbf{e}_{ik}$ was chosen as 5 \si{\angstrom}. The lengths of the node, edge, and global state feature vectors were set at 64, 64, and 16, respectively. The node feature vector is a learned embedding based on the atomic number of each atom, while the global state feature vector is a learned embedding based on the fidelity of the data. The bond angles were expanded using a spherical harmonic basis with $m=3$ and $l=0$. The number of message-passing blocks (comprising sequential three-body interactions, bond, atom and global state updates) is set to 3. The numbers of training parameters in the 1fi and mfi M3GNet models for silicon are 221,597 and 325,757, respectively, while those for water are 221,661 and 325,821, respectively. Each dense layer contains 64 output neurons that are fully connected from the input neurons. The update functions are modeled using gated multi-layer perceptrons (MLPs) that contain two layers with 64 neurons for each layer.
Each updated atomic feature vector is fed into a 64-64-1 gated MLP to predict the corresponding atomic energy.   
The total energy is expressed as the sum of the atomic energies, while the forces and stresses are calculated by taking the partial derivatives of the total energy with respect to the atomic positions and lattice vectors.

\subsection{Potential Training}
All model parameters were optimized using the Adam algorithm\cite{kingma2014adam}. The initial learning rate was set at 0.001. The cosine scheduler was used to gradually reduce the learning rate to 0.01 of the original value in 100 epochs. Early stopping of the model training was triggered when the validation loss did not reduce for 200 epochs. A batch size of 16 and 8 was used in the model training for silicon and water, respectively. However, a smaller batch size of 4 was used for mfi M3GNet training for water to avoid training crashes caused by excessive memory assumption. The loss function for the potential training was given by
\begin{equation}
    L = \mathrm{MSE}(\frac{E_\mathrm{total}^{\mathrm{M3GNet}}}{N_{\mathrm{atoms}}}, \frac{E_\mathrm{total}^{\mathrm{DFT}}}{N_{\mathrm{atoms}}}) + w_{f}\mathrm{MSE}(F_{i,\alpha}^{\mathrm{M3GNet}},F_{i, \alpha}^{\mathrm{DFT}}),  
\end{equation}
where $\mathrm{MSE}$ represents the mean squared error loss function, $E_\mathrm{total}$ denotes the total energy of the system, $N_{\mathrm{atoms}}$ denotes the number of atoms, and $F_{i,\alpha}$ denotes the force component acting on atom $i$ along the $\alpha=x,y,z$ axis. 

$E_f$ is obtained by subtracting the elemental reference energies $\mathbf{E_{\mathrm{elem}}}$ from the total energy, via the following expression:
\begin{equation}
    E_i = E_{\mathrm{total}} - \mathbf{n_{elem}} \cdot \mathbf{E_{\mathrm{elem}}}
\end{equation}
where $\mathbf{n_{elem}}$ is a row vector representing the number of each element in the structure. The elemental reference energies $\mathbf{E_{\mathrm{elem}}}$ were fitted using linear regression of the target total energies, using the following expression:
\begin{equation}
    \mathbf{E_{\mathrm{elem}}} = (\mathbf{A}^{T}\mathbf{A})^{-1}\mathbf{A}^{T}\mathbf{E_{\mathrm{total}}},
\end{equation}
where $\mathbf{E_{\mathrm{total}}}$ is the matrix of total energies with dimensions $N_{\mathrm{struct}} \times N_{\mathrm{elem}}$, $N_{\mathrm{struct}}$ is the total number of structures, $N_{\mathrm{elem}}$ is the total number of elements, and $\mathrm{A}$ is the composition matrix obtained by stacking $\mathbf{n_{elem}}$ for all structures. 

To recover the total energy from the atomic energies, the following expression is used:
\begin{equation}
    E_{total} = \sum_{i}^{N_{\mathrm{atom}}} \sigma E_{i}+\mathbf{n_{elem}} \cdot \mathbf{E_{\mathrm{elem}}}
\end{equation}
where the scaling factor $\sigma$ is calculated by taking the inverse of the root mean square of all atomic force components from the training set.

\subsection{DFT static calculations}

All DFT calculations were performed using VASP (version: 6.3.2) with spin polarization. The projector augmented wave method\cite{blochl1994projector} was employed to describe core and valence electrons with pseudopotentials. The electronic convergence criterion was set at 10$^{-5}$ eV. For the Murnaghan equation of state for silicon crystals, the energy cut-off and k-spacing were chosen to be 800 eV and 0.2 \si{\angstrom}$^{-1}$. 

\subsubsection{Geometry Relaxation}

For the high-throughput geometry relaxation of ice crystals with DFT, the ionic and electronic convergence were set to 0.05 eV ~\si{\angstrom}$^{-1}$ and 10$^{-4}$ eV, respectively. The conjugate gradient algorithm\cite{fletcher1964function} was employed to update the ionic positions. Moreover, the energy cutoff and k-spacing were chosen to be 0.35~\si{\angstrom}$^{-1}$, respectively. 
The same force threshold was applied to the M3GNet geometry relaxation via ASE interface\cite{larsen2017atomic} using the LBFGS\cite{nocedal1980updating} algorithm.

\subsection{MD Simulations}
All MD simulations were performed with LAMMPS\cite{plimpton1995fast,thompson2022lammps} (version: 2Jun2022). A Nose-Hoover thermostat was employed to control temperature and pressure for \textit{NPT} simulations. The time step was chosen to be 1 fs and the damping constants for temperature and pressure were set to 0.1 ps and 1.0 ps, respectively. 

Amorphous silicon structures were generated by the melt-and-quench approach. The initial structure was obtained by random displacement of 1728 silicon atoms in the cubic cell with a density of 2.56 g/cm$^{3}$ and then relaxed under a force threshold of 10$^{-3}$ eV \si{\angstrom}$^{-1}$. The relaxed structure was equilibrated at 1500 K for 110 ps and then cooled down to 500 K. A fast quenching rate of 10$^{13}$K s$^{-1}$ was used for quenching from 1500 K to 1250 K and from 1050K to 500K, while a slower quenching rate of 10$^{11}$K s$^{-1}$ was used for quenching from 1250 K and 1050 K. The simulated temperature and the corresponding potential energy of silicon during the melt-and-quench are provided in Fig. S3.
After an additional 50 ps equilibration at 500K, the amorphous silicon was isotropically compressed at a rate of 0.5 GPa (ps)$^{-1}$ from 0 GPa to 20GPa. The snapshot of silicon structures at 20 GPa was further equilibrated for 20 ps at 500K. Finally, their structural properties were obtained from 100-ps NVE simulations.
For the \textit{NPT} simulations of liquid water, an initial cubic cell containing 64 H\textsubscript{2}O molecules with $\rho=1.0$ g cm\textsuperscript{-3} was equilibrated for 50 ps under 1 atm. The density and structural properties were extracted from a production run of 0.5 ns.

\subsection{DIRECT Sampling}

The DImensionally-Reduced Encoded Clusters with sTratified (DIRECT) sampling workflow has already been detailed in previous work.\cite{qi2024robust}  First, a pre-trained M3GNet Materials Project formation energy model was used to encode structures into a 128-element vector. Next, principal components analysis (PCA) was performed on the encoded structure features, and the first six PCs with eigenvalues greater than 1 (Kaiser's rule) were selected as a reduced-dimensionality representation. Finally, the balanced iterative reducing and clustering using hierarchies (BIRCH) algorithm\cite{zhang1996birch} was used to divide all encoded structures into \textit{n} clusters and \textit{k} structures are sampled from each cluster for static DFT calculations. In this work, the values of \textit{n}, \textit{k}, and the threshold used to construct the 1fi and mfi M3GNet GLPs of silicon are given in Table S1.

\subsection{Training Data Selection}

For silicon, 80\% of the training structures were selected from a total of 608 structures, with 70\% randomly sampled and 10\% DIRECT sampled to ensure the training set covered mostly diverse structures while keeping sufficient structural diversity for meaningful validation and testing. The hifi SCAN data points for training the 10\%-mfi-narrow M3GNet model were selected based on PCA, where their first and second PC values were lower than -12.5 and -2.5, respectively. As for water, all training structures were randomly sampled for constructing 1fi and mfi M3GNet GLPs.

\subsection{T-SNE Visualization}
The t-Distributed Stochastic Neighbor Embedding (t-SNE) analysis was performed using the OpenTSNE library\cite{polivcar2019opentsne,linderman2019fast}. The cosine distance was used as the distance metric. All other settings were kept at the default values.

\begin{acknowledgement}
This work was intellectually led by the U.S. Department of Energy, Office of Science, Office of Basic Energy Sciences, Materials Sciences and Engineering Division under contract No. DE-AC02-05-CH11231 (Materials Project program KC23MP). This research used resources of the National Energy Research Scientific Computing Center (NERSC), a Department of Energy Office of Science User Facility using NERSC award DOE-ERCAP0026371. T. W. Ko also acknowledges the support of the Eric and Wendy Schmidt AI in Science Postdoctoral Fellowship, a Schmidt Futures program.

\end{acknowledgement}

\section{Data Availability}
The RPBE water dataset is available at \url{https://zenodo.org/records/2634098}, while the SCAN water dataset is available in ref.\cite{yao2020temperature}. Both PBE and SCAN silicon datasets are given in \url{https://doi.org/10.5281/zenodo.4174139}. 
All data necessary to replicate the training results will be publicly available upon acceptance of the manuscript.

\section{Code Availability}
All codes used in this work are available in public Github repositories.
The M3GNet implementation is available at \url{https://github.com/materialsvirtuallab/m3gnet}. The DIRECT sampling is implemented in the MAterials Machine Learning (maml) (\url{https://github.com/materialsvirtuallab/maml}) repository. 

\section{Author contributions}
T. W. Ko and S. P. Ong conceived the idea and designed the work. T. W. Ko modified the original M3GNet implementations and performed the analysis. T. W. Ko and S. P. Ong wrote the manuscript and contributed to the discussion and revision.

\section{Completing Interests}
The authors declare that they have no competing interests.


\providecommand{\latin}[1]{#1}
\makeatletter
\providecommand{\doi}
  {\begingroup\let\do\@makeother\dospecials
  \catcode`\{=1 \catcode`\}=2 \doi@aux}
\providecommand{\doi@aux}[1]{\endgroup\texttt{#1}}
\makeatother
\providecommand*\mcitethebibliography{\thebibliography}
\csname @ifundefined\endcsname{endmcitethebibliography}
  {\let\endmcitethebibliography\endthebibliography}{}
\begin{mcitethebibliography}{61}
\providecommand*\natexlab[1]{#1}
\providecommand*\mciteSetBstSublistMode[1]{}
\providecommand*\mciteSetBstMaxWidthForm[2]{}
\providecommand*\mciteBstWouldAddEndPuncttrue
  {\def\EndOfBibitem{\unskip.}}
\providecommand*\mciteBstWouldAddEndPunctfalse
  {\let\EndOfBibitem\relax}
\providecommand*\mciteSetBstMidEndSepPunct[3]{}
\providecommand*\mciteSetBstSublistLabelBeginEnd[3]{}
\providecommand*\EndOfBibitem{}
\mciteSetBstSublistMode{f}
\mciteSetBstMaxWidthForm{subitem}{(\alph{mcitesubitemcount})}
\mciteSetBstSublistLabelBeginEnd
  {\mcitemaxwidthsubitemform\space}
  {\relax}
  {\relax}

\bibitem[Unke \latin{et~al.}(2021)Unke, Chmiela, Sauceda, Gastegger, Poltavsky,
  Sch{\"u}tt, Tkatchenko, and M{\"u}ller]{unke2021machine}
Unke,~O.~T.; Chmiela,~S.; Sauceda,~H.~E.; Gastegger,~M.; Poltavsky,~I.;
  Sch{\"u}tt,~K.~T.; Tkatchenko,~A.; M{\"u}ller,~K.-R. Machine learning force
  fields. \emph{Chem. Rev.} \textbf{2021}, \emph{121}, 10142--10186\relax
\mciteBstWouldAddEndPuncttrue
\mciteSetBstMidEndSepPunct{\mcitedefaultmidpunct}
{\mcitedefaultendpunct}{\mcitedefaultseppunct}\relax
\EndOfBibitem
\bibitem[Ko \latin{et~al.}(2021)Ko, Finkler, Goedecker, and
  Behler]{ko2021general}
Ko,~T.~W.; Finkler,~J.~A.; Goedecker,~S.; Behler,~J. General-purpose machine
  learning potentials capturing nonlocal charge transfer. \emph{Acc. Chem.
  Res.} \textbf{2021}, \emph{54}, 808--817\relax
\mciteBstWouldAddEndPuncttrue
\mciteSetBstMidEndSepPunct{\mcitedefaultmidpunct}
{\mcitedefaultendpunct}{\mcitedefaultseppunct}\relax
\EndOfBibitem
\bibitem[Kocer \latin{et~al.}(2022)Kocer, Ko, and Behler]{kocer2022neural}
Kocer,~E.; Ko,~T.~W.; Behler,~J. Neural network potentials: A concise overview
  of methods. \emph{Ann. Rev. Phys. Chem.} \textbf{2022}, \emph{73},
  163--186\relax
\mciteBstWouldAddEndPuncttrue
\mciteSetBstMidEndSepPunct{\mcitedefaultmidpunct}
{\mcitedefaultendpunct}{\mcitedefaultseppunct}\relax
\EndOfBibitem
\bibitem[Behler and Parrinello(2007)Behler, and
  Parrinello]{behler2007generalized}
Behler,~J.; Parrinello,~M. Generalized neural-network representation of
  high-dimensional potential-energy surfaces. \emph{Phys. Rev. Lett.}
  \textbf{2007}, \emph{98}, 146401\relax
\mciteBstWouldAddEndPuncttrue
\mciteSetBstMidEndSepPunct{\mcitedefaultmidpunct}
{\mcitedefaultendpunct}{\mcitedefaultseppunct}\relax
\EndOfBibitem
\bibitem[Ko \latin{et~al.}(2021)Ko, Finkler, Goedecker, and
  Behler]{ko2021fourth}
Ko,~T.~W.; Finkler,~J.~A.; Goedecker,~S.; Behler,~J. A fourth-generation
  high-dimensional neural network potential with accurate electrostatics
  including non-local charge transfer. \emph{Nat. Commun.} \textbf{2021},
  \emph{12}, 398\relax
\mciteBstWouldAddEndPuncttrue
\mciteSetBstMidEndSepPunct{\mcitedefaultmidpunct}
{\mcitedefaultendpunct}{\mcitedefaultseppunct}\relax
\EndOfBibitem
\bibitem[Ko \latin{et~al.}(2023)Ko, Finkler, Goedecker, and
  Behler]{ko2023accurate}
Ko,~T.~W.; Finkler,~J.~A.; Goedecker,~S.; Behler,~J. Accurate Fourth-Generation
  Machine Learning Potentials by Electrostatic Embedding. \emph{J. Chem. Theory
  Comput.} \textbf{2023}, \relax
\mciteBstWouldAddEndPunctfalse
\mciteSetBstMidEndSepPunct{\mcitedefaultmidpunct}
{}{\mcitedefaultseppunct}\relax
\EndOfBibitem
\bibitem[Bart{\'o}k \latin{et~al.}(2010)Bart{\'o}k, Payne, Kondor, and
  Cs{\'a}nyi]{bartok2010gaussian}
Bart{\'o}k,~A.~P.; Payne,~M.~C.; Kondor,~R.; Cs{\'a}nyi,~G. Gaussian
  approximation potentials: The accuracy of quantum mechanics, without the
  electrons. \emph{Phys. Rev. Lett.} \textbf{2010}, \emph{104}, 136403\relax
\mciteBstWouldAddEndPuncttrue
\mciteSetBstMidEndSepPunct{\mcitedefaultmidpunct}
{\mcitedefaultendpunct}{\mcitedefaultseppunct}\relax
\EndOfBibitem
\bibitem[Thompson \latin{et~al.}(2015)Thompson, Swiler, Trott, Foiles, and
  Tucker]{thompson2015spectral}
Thompson,~A.~P.; Swiler,~L.~P.; Trott,~C.~R.; Foiles,~S.~M.; Tucker,~G.~J.
  Spectral neighbor analysis method for automated generation of
  quantum-accurate interatomic potentials. \emph{J. Comput. Phys.}
  \textbf{2015}, \emph{285}, 316--330\relax
\mciteBstWouldAddEndPuncttrue
\mciteSetBstMidEndSepPunct{\mcitedefaultmidpunct}
{\mcitedefaultendpunct}{\mcitedefaultseppunct}\relax
\EndOfBibitem
\bibitem[Shapeev(2016)]{shapeev2016moment}
Shapeev,~A.~V. Moment tensor potentials: A class of systematically improvable
  interatomic potentials. \emph{Multiscale Model. Simul.} \textbf{2016},
  \emph{14}, 1153--1173\relax
\mciteBstWouldAddEndPuncttrue
\mciteSetBstMidEndSepPunct{\mcitedefaultmidpunct}
{\mcitedefaultendpunct}{\mcitedefaultseppunct}\relax
\EndOfBibitem
\bibitem[Drautz(2019)]{drautz2019atomic}
Drautz,~R. Atomic cluster expansion for accurate and transferable interatomic
  potentials. \emph{Phys. Rev. B} \textbf{2019}, \emph{99}, 014104\relax
\mciteBstWouldAddEndPuncttrue
\mciteSetBstMidEndSepPunct{\mcitedefaultmidpunct}
{\mcitedefaultendpunct}{\mcitedefaultseppunct}\relax
\EndOfBibitem
\bibitem[Zuo \latin{et~al.}(2020)Zuo, Chen, Li, Deng, Chen, Behler, Cs{\'a}nyi,
  Shapeev, Thompson, Wood, \latin{et~al.} others]{zuo2020performance}
Zuo,~Y.; Chen,~C.; Li,~X.; Deng,~Z.; Chen,~Y.; Behler,~J.; Cs{\'a}nyi,~G.;
  Shapeev,~A.~V.; Thompson,~A.~P.; Wood,~M.~A.; others Performance and cost
  assessment of machine learning interatomic potentials. \emph{J. Phys. Chem.
  A} \textbf{2020}, \emph{124}, 731--745\relax
\mciteBstWouldAddEndPuncttrue
\mciteSetBstMidEndSepPunct{\mcitedefaultmidpunct}
{\mcitedefaultendpunct}{\mcitedefaultseppunct}\relax
\EndOfBibitem
\bibitem[Li \latin{et~al.}(2020)Li, Chen, Zheng, Zuo, and Ong]{li2020complex}
Li,~X.-G.; Chen,~C.; Zheng,~H.; Zuo,~Y.; Ong,~S.~P. Complex strengthening
  mechanisms in the NbMoTaW multi-principal element alloy. \emph{npj Comput.
  Mater.} \textbf{2020}, \emph{6}, 70\relax
\mciteBstWouldAddEndPuncttrue
\mciteSetBstMidEndSepPunct{\mcitedefaultmidpunct}
{\mcitedefaultendpunct}{\mcitedefaultseppunct}\relax
\EndOfBibitem
\bibitem[Lee \latin{et~al.}(2023)Lee, Qi, Gadre, Huyan, Ko, Zuo, Du, Li, Aoki,
  Wu, \latin{et~al.} others]{lee2023atomic}
Lee,~T.; Qi,~J.; Gadre,~C.~A.; Huyan,~H.; Ko,~S.-T.; Zuo,~Y.; Du,~C.; Li,~J.;
  Aoki,~T.; Wu,~R.; others Atomic-scale origin of the low grain-boundary
  resistance in perovskite solid electrolyte Li0. 375Sr0. 4375Ta0. 75Zr0. 25O3.
  \emph{Nat. Commun.} \textbf{2023}, \emph{14}, 1940\relax
\mciteBstWouldAddEndPuncttrue
\mciteSetBstMidEndSepPunct{\mcitedefaultmidpunct}
{\mcitedefaultendpunct}{\mcitedefaultseppunct}\relax
\EndOfBibitem
\bibitem[Kostiuchenko \latin{et~al.}(2019)Kostiuchenko, K{\"o}rmann,
  Neugebauer, and Shapeev]{kostiuchenko2019impact}
Kostiuchenko,~T.; K{\"o}rmann,~F.; Neugebauer,~J.; Shapeev,~A. Impact of
  lattice relaxations on phase transitions in a high-entropy alloy studied by
  machine-learning potentials. \emph{npj Comput. Mater.} \textbf{2019},
  \emph{5}, 55\relax
\mciteBstWouldAddEndPuncttrue
\mciteSetBstMidEndSepPunct{\mcitedefaultmidpunct}
{\mcitedefaultendpunct}{\mcitedefaultseppunct}\relax
\EndOfBibitem
\bibitem[Santos-Florez \latin{et~al.}(2023)Santos-Florez, Dai, Yao, Yanxon, Li,
  Wang, Zhu, and Yu]{santos2023short}
Santos-Florez,~P.~A.; Dai,~S.-C.; Yao,~Y.; Yanxon,~H.; Li,~L.; Wang,~Y.-J.;
  Zhu,~Q.; Yu,~X.-X. Short-range order and its impacts on the BCC MoNbTaW
  multi-principal element alloy by the machine-learning potential. \emph{Acta
  Mater.} \textbf{2023}, \emph{255}, 119041\relax
\mciteBstWouldAddEndPuncttrue
\mciteSetBstMidEndSepPunct{\mcitedefaultmidpunct}
{\mcitedefaultendpunct}{\mcitedefaultseppunct}\relax
\EndOfBibitem
\bibitem[Byggm{\"a}star \latin{et~al.}(2021)Byggm{\"a}star, Nordlund, and
  Djurabekova]{byggmastar2021modeling}
Byggm{\"a}star,~J.; Nordlund,~K.; Djurabekova,~F. Modeling refractory
  high-entropy alloys with efficient machine-learned interatomic potentials:
  Defects and segregation. \emph{Phys. Rev. B} \textbf{2021}, \emph{104},
  104101\relax
\mciteBstWouldAddEndPuncttrue
\mciteSetBstMidEndSepPunct{\mcitedefaultmidpunct}
{\mcitedefaultendpunct}{\mcitedefaultseppunct}\relax
\EndOfBibitem
\bibitem[Qi \latin{et~al.}(2021)Qi, Banerjee, Zuo, Chen, Zhu, Chandrappa, Li,
  and Ong]{qi2021bridging}
Qi,~J.; Banerjee,~S.; Zuo,~Y.; Chen,~C.; Zhu,~Z.; Chandrappa,~M.~H.; Li,~X.;
  Ong,~S.~P. Bridging the gap between simulated and experimental ionic
  conductivities in lithium superionic conductors. \emph{Mater. Today Phys.}
  \textbf{2021}, \emph{21}, 100463\relax
\mciteBstWouldAddEndPuncttrue
\mciteSetBstMidEndSepPunct{\mcitedefaultmidpunct}
{\mcitedefaultendpunct}{\mcitedefaultseppunct}\relax
\EndOfBibitem
\bibitem[Krenzer \latin{et~al.}(2023)Krenzer, Klarbring, Tolborg, Rossignol,
  McCluskey, Morgan, and Walsh]{krenzer2023nature}
Krenzer,~G.; Klarbring,~J.; Tolborg,~K.; Rossignol,~H.; McCluskey,~A.~R.;
  Morgan,~B.~J.; Walsh,~A. Nature of the superionic phase transition of lithium
  nitride from machine learning force fields. \emph{Chem. Mater.}
  \textbf{2023}, \emph{35}, 6133--6140\relax
\mciteBstWouldAddEndPuncttrue
\mciteSetBstMidEndSepPunct{\mcitedefaultmidpunct}
{\mcitedefaultendpunct}{\mcitedefaultseppunct}\relax
\EndOfBibitem
\bibitem[Lacivita \latin{et~al.}(2018)Lacivita, Artrith, and
  Ceder]{lacivita2018structural}
Lacivita,~V.; Artrith,~N.; Ceder,~G. Structural and compositional factors that
  control the Li-ion conductivity in LiPON electrolytes. \emph{Chem. Mater.}
  \textbf{2018}, \emph{30}, 7077--7090\relax
\mciteBstWouldAddEndPuncttrue
\mciteSetBstMidEndSepPunct{\mcitedefaultmidpunct}
{\mcitedefaultendpunct}{\mcitedefaultseppunct}\relax
\EndOfBibitem
\bibitem[Chen and Ong(2022)Chen, and Ong]{chen2022universal}
Chen,~C.; Ong,~S.~P. A universal graph deep learning interatomic potential for
  the periodic table. \emph{Nat. Comput. Sci.} \textbf{2022}, \emph{2},
  718--728\relax
\mciteBstWouldAddEndPuncttrue
\mciteSetBstMidEndSepPunct{\mcitedefaultmidpunct}
{\mcitedefaultendpunct}{\mcitedefaultseppunct}\relax
\EndOfBibitem
\bibitem[Deng \latin{et~al.}(2023)Deng, Zhong, Jun, Riebesell, Han, Bartel, and
  Ceder]{deng2023chgnet}
Deng,~B.; Zhong,~P.; Jun,~K.; Riebesell,~J.; Han,~K.; Bartel,~C.~J.; Ceder,~G.
  CHGNet as a pretrained universal neural network potential for charge-informed
  atomistic modelling. \emph{Nat. Mach. Intell.} \textbf{2023}, \emph{5},
  1031--1041\relax
\mciteBstWouldAddEndPuncttrue
\mciteSetBstMidEndSepPunct{\mcitedefaultmidpunct}
{\mcitedefaultendpunct}{\mcitedefaultseppunct}\relax
\EndOfBibitem
\bibitem[Batatia \latin{et~al.}(2022)Batatia, Kovacs, Simm, Ortner, and
  Cs{\'a}nyi]{batatia2022mace}
Batatia,~I.; Kovacs,~D.~P.; Simm,~G.; Ortner,~C.; Cs{\'a}nyi,~G. MACE: Higher
  order equivariant message passing neural networks for fast and accurate force
  fields. \emph{Adv. Neural Inf. Process. Syst.} \textbf{2022}, \emph{35},
  11423--11436\relax
\mciteBstWouldAddEndPuncttrue
\mciteSetBstMidEndSepPunct{\mcitedefaultmidpunct}
{\mcitedefaultendpunct}{\mcitedefaultseppunct}\relax
\EndOfBibitem
\bibitem[Batzner \latin{et~al.}(2022)Batzner, Musaelian, Sun, Geiger, Mailoa,
  Kornbluth, Molinari, Smidt, and Kozinsky]{batzner20223}
Batzner,~S.; Musaelian,~A.; Sun,~L.; Geiger,~M.; Mailoa,~J.~P.; Kornbluth,~M.;
  Molinari,~N.; Smidt,~T.~E.; Kozinsky,~B. E (3)-equivariant graph neural
  networks for data-efficient and accurate interatomic potentials. \emph{Nat.
  Commun.} \textbf{2022}, \emph{13}, 2453\relax
\mciteBstWouldAddEndPuncttrue
\mciteSetBstMidEndSepPunct{\mcitedefaultmidpunct}
{\mcitedefaultendpunct}{\mcitedefaultseppunct}\relax
\EndOfBibitem
\bibitem[Ko and Ong(2023)Ko, and Ong]{ko2023recent}
Ko,~T.~W.; Ong,~S.~P. Recent advances and outstanding challenges for machine
  learning interatomic potentials. \emph{Nat. Comput. Sci.} \textbf{2023},
  1--3\relax
\mciteBstWouldAddEndPuncttrue
\mciteSetBstMidEndSepPunct{\mcitedefaultmidpunct}
{\mcitedefaultendpunct}{\mcitedefaultseppunct}\relax
\EndOfBibitem
\bibitem[Jain \latin{et~al.}(2013)Jain, Ong, Hautier, Chen, Richards, Dacek,
  Cholia, Gunter, Skinner, Ceder, \latin{et~al.} others]{jain2013commentary}
Jain,~A.; Ong,~S.~P.; Hautier,~G.; Chen,~W.; Richards,~W.~D.; Dacek,~S.;
  Cholia,~S.; Gunter,~D.; Skinner,~D.; Ceder,~G.; others Commentary: The
  Materials Project: A materials genome approach to accelerating materials
  innovation. \emph{APL Mater.} \textbf{2013}, \emph{1}\relax
\mciteBstWouldAddEndPuncttrue
\mciteSetBstMidEndSepPunct{\mcitedefaultmidpunct}
{\mcitedefaultendpunct}{\mcitedefaultseppunct}\relax
\EndOfBibitem
\bibitem[Perdew \latin{et~al.}(1996)Perdew, Burke, and
  Ernzerhof]{perdew1996generalized}
Perdew,~J.~P.; Burke,~K.; Ernzerhof,~M. Generalized gradient approximation made
  simple. \emph{Phys. Rev. Lett.} \textbf{1996}, \emph{77}, 3865\relax
\mciteBstWouldAddEndPuncttrue
\mciteSetBstMidEndSepPunct{\mcitedefaultmidpunct}
{\mcitedefaultendpunct}{\mcitedefaultseppunct}\relax
\EndOfBibitem
\bibitem[Sun \latin{et~al.}(2015)Sun, Ruzsinszky, and Perdew]{sun2015strongly}
Sun,~J.; Ruzsinszky,~A.; Perdew,~J.~P. Strongly constrained and appropriately
  normed semilocal density functional. \emph{Phys. Rev. Lett.} \textbf{2015},
  \emph{115}, 036402\relax
\mciteBstWouldAddEndPuncttrue
\mciteSetBstMidEndSepPunct{\mcitedefaultmidpunct}
{\mcitedefaultendpunct}{\mcitedefaultseppunct}\relax
\EndOfBibitem
\bibitem[Furness \latin{et~al.}(2020)Furness, Kaplan, Ning, Perdew, and
  Sun]{furness2020accurate}
Furness,~J.~W.; Kaplan,~A.~D.; Ning,~J.; Perdew,~J.~P.; Sun,~J. Accurate and
  numerically efficient r2SCAN meta-generalized gradient approximation.
  \emph{J. Phys. Chem. Lett.} \textbf{2020}, \emph{11}, 8208--8215\relax
\mciteBstWouldAddEndPuncttrue
\mciteSetBstMidEndSepPunct{\mcitedefaultmidpunct}
{\mcitedefaultendpunct}{\mcitedefaultseppunct}\relax
\EndOfBibitem
\bibitem[Chen \latin{et~al.}(2021)Chen, Zuo, Ye, Li, and Ong]{chen2021learning}
Chen,~C.; Zuo,~Y.; Ye,~W.; Li,~X.; Ong,~S.~P. Learning properties of ordered
  and disordered materials from multi-fidelity data. \emph{Nat. Comput. Sci.}
  \textbf{2021}, \emph{1}, 46--53\relax
\mciteBstWouldAddEndPuncttrue
\mciteSetBstMidEndSepPunct{\mcitedefaultmidpunct}
{\mcitedefaultendpunct}{\mcitedefaultseppunct}\relax
\EndOfBibitem
\bibitem[Cooper \latin{et~al.}(2000)Cooper, Goringe, and
  McKenzie]{cooper2000density}
Cooper,~N.; Goringe,~C.; McKenzie,~D. Density functional theory modelling of
  amorphous silicon. \emph{Comput. Mater. Sci.} \textbf{2000}, \emph{17},
  1--6\relax
\mciteBstWouldAddEndPuncttrue
\mciteSetBstMidEndSepPunct{\mcitedefaultmidpunct}
{\mcitedefaultendpunct}{\mcitedefaultseppunct}\relax
\EndOfBibitem
\bibitem[Pedersen \latin{et~al.}(2017)Pedersen, Pizzagalli, and
  J{\'o}nsson]{pedersen2017optimal}
Pedersen,~A.; Pizzagalli,~L.; J{\'o}nsson,~H. Optimal atomic structure of
  amorphous silicon obtained from density functional theory calculations.
  \emph{New J. Phys.} \textbf{2017}, \emph{19}, 063018\relax
\mciteBstWouldAddEndPuncttrue
\mciteSetBstMidEndSepPunct{\mcitedefaultmidpunct}
{\mcitedefaultendpunct}{\mcitedefaultseppunct}\relax
\EndOfBibitem
\bibitem[Remsing \latin{et~al.}(2017)Remsing, Klein, and
  Sun]{remsing2017dependence}
Remsing,~R.~C.; Klein,~M.~L.; Sun,~J. Dependence of the structure and dynamics
  of liquid silicon on the choice of density functional approximation.
  \emph{Phys. Rev. B} \textbf{2017}, \emph{96}, 024203\relax
\mciteBstWouldAddEndPuncttrue
\mciteSetBstMidEndSepPunct{\mcitedefaultmidpunct}
{\mcitedefaultendpunct}{\mcitedefaultseppunct}\relax
\EndOfBibitem
\bibitem[Remsing \latin{et~al.}(2018)Remsing, Klein, and
  Sun]{remsing2018refined}
Remsing,~R.~C.; Klein,~M.~L.; Sun,~J. Refined description of liquid and
  supercooled silicon from ab initio simulations. \emph{Phys. Rev. B}
  \textbf{2018}, \emph{97}, 140103\relax
\mciteBstWouldAddEndPuncttrue
\mciteSetBstMidEndSepPunct{\mcitedefaultmidpunct}
{\mcitedefaultendpunct}{\mcitedefaultseppunct}\relax
\EndOfBibitem
\bibitem[Cheng \latin{et~al.}(2019)Cheng, Engel, Behler, Dellago, and
  Ceriotti]{cheng2019ab}
Cheng,~B.; Engel,~E.~A.; Behler,~J.; Dellago,~C.; Ceriotti,~M. Ab initio
  thermodynamics of liquid and solid water. \emph{Proc. Natl. Acad. Sci.}
  \textbf{2019}, \emph{116}, 1110--1115\relax
\mciteBstWouldAddEndPuncttrue
\mciteSetBstMidEndSepPunct{\mcitedefaultmidpunct}
{\mcitedefaultendpunct}{\mcitedefaultseppunct}\relax
\EndOfBibitem
\bibitem[Ruiz~Pestana \latin{et~al.}(2018)Ruiz~Pestana, Marsalek, Markland, and
  Head-Gordon]{ruiz2018quest}
Ruiz~Pestana,~L.; Marsalek,~O.; Markland,~T.~E.; Head-Gordon,~T. The quest for
  accurate liquid water properties from first principles. \emph{J. Phys. Chem.
  Lett.} \textbf{2018}, \emph{9}, 5009--5016\relax
\mciteBstWouldAddEndPuncttrue
\mciteSetBstMidEndSepPunct{\mcitedefaultmidpunct}
{\mcitedefaultendpunct}{\mcitedefaultseppunct}\relax
\EndOfBibitem
\bibitem[Forster-Tonigold and Groß(2014)Forster-Tonigold, and
  Groß]{forster2014dispersion}
Forster-Tonigold,~K.; Groß,~A. Dispersion corrected RPBE studies of liquid
  water. \emph{J. Chem. Phys.} \textbf{2014}, \emph{141}\relax
\mciteBstWouldAddEndPuncttrue
\mciteSetBstMidEndSepPunct{\mcitedefaultmidpunct}
{\mcitedefaultendpunct}{\mcitedefaultseppunct}\relax
\EndOfBibitem
\bibitem[Sun \latin{et~al.}(2016)Sun, Remsing, Zhang, Sun, Ruzsinszky, Peng,
  Yang, Paul, Waghmare, Wu, \latin{et~al.} others]{sun2016accurate}
Sun,~J.; Remsing,~R.~C.; Zhang,~Y.; Sun,~Z.; Ruzsinszky,~A.; Peng,~H.;
  Yang,~Z.; Paul,~A.; Waghmare,~U.; Wu,~X.; others Accurate first-principles
  structures and energies of diversely bonded systems from an efficient density
  functional. \emph{Nat. Chem.} \textbf{2016}, \emph{8}, 831--836\relax
\mciteBstWouldAddEndPuncttrue
\mciteSetBstMidEndSepPunct{\mcitedefaultmidpunct}
{\mcitedefaultendpunct}{\mcitedefaultseppunct}\relax
\EndOfBibitem
\bibitem[Deringer \latin{et~al.}(2021)Deringer, Bernstein, Cs{\'a}nyi,
  Ben~Mahmoud, Ceriotti, Wilson, Drabold, and Elliott]{deringer2021origins}
Deringer,~V.~L.; Bernstein,~N.; Cs{\'a}nyi,~G.; Ben~Mahmoud,~C.; Ceriotti,~M.;
  Wilson,~M.; Drabold,~D.~A.; Elliott,~S.~R. Origins of structural and
  electronic transitions in disordered silicon. \emph{Nature} \textbf{2021},
  \emph{589}, 59--64\relax
\mciteBstWouldAddEndPuncttrue
\mciteSetBstMidEndSepPunct{\mcitedefaultmidpunct}
{\mcitedefaultendpunct}{\mcitedefaultseppunct}\relax
\EndOfBibitem
\bibitem[Bart{\'o}k \latin{et~al.}(2018)Bart{\'o}k, Kermode, Bernstein, and
  Cs{\'a}nyi]{bartok2018machine}
Bart{\'o}k,~A.~P.; Kermode,~J.; Bernstein,~N.; Cs{\'a}nyi,~G. Machine learning
  a general-purpose interatomic potential for silicon. \emph{Phys. Rev. X}
  \textbf{2018}, \emph{8}, 041048\relax
\mciteBstWouldAddEndPuncttrue
\mciteSetBstMidEndSepPunct{\mcitedefaultmidpunct}
{\mcitedefaultendpunct}{\mcitedefaultseppunct}\relax
\EndOfBibitem
\bibitem[Zheng \latin{et~al.}(2018)Zheng, Chen, Sun, Ko, Santra, Dhuvad, and
  Wu]{zheng2018structural}
Zheng,~L.; Chen,~M.; Sun,~Z.; Ko,~H.-Y.; Santra,~B.; Dhuvad,~P.; Wu,~X.
  Structural, electronic, and dynamical properties of liquid water by ab initio
  molecular dynamics based on SCAN functional within the canonical ensemble.
  \emph{J. Chem. Phys.} \textbf{2018}, \emph{148}\relax
\mciteBstWouldAddEndPuncttrue
\mciteSetBstMidEndSepPunct{\mcitedefaultmidpunct}
{\mcitedefaultendpunct}{\mcitedefaultseppunct}\relax
\EndOfBibitem
\bibitem[Chen \latin{et~al.}(2017)Chen, Ko, Remsing, Calegari~Andrade, Santra,
  Sun, Selloni, Car, Klein, Perdew, \latin{et~al.} others]{chen2017ab}
Chen,~M.; Ko,~H.-Y.; Remsing,~R.~C.; Calegari~Andrade,~M.~F.; Santra,~B.;
  Sun,~Z.; Selloni,~A.; Car,~R.; Klein,~M.~L.; Perdew,~J.~P.; others Ab initio
  theory and modeling of water. \emph{Proc. Natl. Acad. Sci.} \textbf{2017},
  \emph{114}, 10846--10851\relax
\mciteBstWouldAddEndPuncttrue
\mciteSetBstMidEndSepPunct{\mcitedefaultmidpunct}
{\mcitedefaultendpunct}{\mcitedefaultseppunct}\relax
\EndOfBibitem
\bibitem[Zhang \latin{et~al.}(2021)Zhang, Tang, Chen, Xu, Zhang, Qiu, Perdew,
  Klein, and Wu]{zhang2021modeling}
Zhang,~C.; Tang,~F.; Chen,~M.; Xu,~J.; Zhang,~L.; Qiu,~D.~Y.; Perdew,~J.~P.;
  Klein,~M.~L.; Wu,~X. Modeling liquid water by climbing up Jacob’s ladder in
  density functional theory facilitated by using deep neural network
  potentials. \emph{J. Phys. Chem. B} \textbf{2021}, \emph{125},
  11444--11456\relax
\mciteBstWouldAddEndPuncttrue
\mciteSetBstMidEndSepPunct{\mcitedefaultmidpunct}
{\mcitedefaultendpunct}{\mcitedefaultseppunct}\relax
\EndOfBibitem
\bibitem[Morawietz \latin{et~al.}(2016)Morawietz, Singraber, Dellago, and
  Behler]{morawietz2016van}
Morawietz,~T.; Singraber,~A.; Dellago,~C.; Behler,~J. How van der Waals
  interactions determine the unique properties of water. \emph{Proc. Natl.
  Acad. Sci.} \textbf{2016}, \emph{113}, 8368--8373\relax
\mciteBstWouldAddEndPuncttrue
\mciteSetBstMidEndSepPunct{\mcitedefaultmidpunct}
{\mcitedefaultendpunct}{\mcitedefaultseppunct}\relax
\EndOfBibitem
\bibitem[Sprik \latin{et~al.}(1996)Sprik, Hutter, and Parrinello]{sprik1996ab}
Sprik,~M.; Hutter,~J.; Parrinello,~M. Ab initio molecular dynamics simulation
  of liquid water: Comparison of three gradient-corrected density functionals.
  \emph{J. Chem. Phys} \textbf{1996}, \emph{105}, 1142--1152\relax
\mciteBstWouldAddEndPuncttrue
\mciteSetBstMidEndSepPunct{\mcitedefaultmidpunct}
{\mcitedefaultendpunct}{\mcitedefaultseppunct}\relax
\EndOfBibitem
\bibitem[Fern{\'a}ndez-Serra \latin{et~al.}(2005)Fern{\'a}ndez-Serra, Ferlat,
  and Artacho]{fernandez2005two}
Fern{\'a}ndez-Serra,~M.; Ferlat,~G.; Artacho,~E. Two exchange-correlation
  functionals compared for first-principles liquid water. \emph{Mol. Simul.}
  \textbf{2005}, \emph{31}, 361--366\relax
\mciteBstWouldAddEndPuncttrue
\mciteSetBstMidEndSepPunct{\mcitedefaultmidpunct}
{\mcitedefaultendpunct}{\mcitedefaultseppunct}\relax
\EndOfBibitem
\bibitem[Stukowski(2009)]{stukowski2009visualization}
Stukowski,~A. Visualization and analysis of atomistic simulation data with
  OVITO--the Open Visualization Tool. \emph{Model. Simul. Mat. Sci. Eng.}
  \textbf{2009}, \emph{18}, 015012\relax
\mciteBstWouldAddEndPuncttrue
\mciteSetBstMidEndSepPunct{\mcitedefaultmidpunct}
{\mcitedefaultendpunct}{\mcitedefaultseppunct}\relax
\EndOfBibitem
\bibitem[Yao and Kanai(2020)Yao, and Kanai]{yao2020temperature}
Yao,~Y.; Kanai,~Y. Temperature dependence of nuclear quantum effects on liquid
  water via artificial neural network model based on SCAN meta-GGA functional.
  \emph{J. Chem. Phys.} \textbf{2020}, \emph{153}\relax
\mciteBstWouldAddEndPuncttrue
\mciteSetBstMidEndSepPunct{\mcitedefaultmidpunct}
{\mcitedefaultendpunct}{\mcitedefaultseppunct}\relax
\EndOfBibitem
\bibitem[Monserrat \latin{et~al.}(2020)Monserrat, Brandenburg, Engel, and
  Cheng]{monserrat2020liquid}
Monserrat,~B.; Brandenburg,~J.~G.; Engel,~E.~A.; Cheng,~B. Liquid water
  contains the building blocks of diverse ice phases. \emph{Nat. Commun.}
  \textbf{2020}, \emph{11}, 5757\relax
\mciteBstWouldAddEndPuncttrue
\mciteSetBstMidEndSepPunct{\mcitedefaultmidpunct}
{\mcitedefaultendpunct}{\mcitedefaultseppunct}\relax
\EndOfBibitem
\bibitem[Pan \latin{et~al.}(2021)Pan, Ganose, Horton, Aykol, Persson,
  Zimmermann, and Jain]{pan2021benchmarking}
Pan,~H.; Ganose,~A.~M.; Horton,~M.; Aykol,~M.; Persson,~K.~A.;
  Zimmermann,~N.~E.; Jain,~A. Benchmarking coordination number prediction
  algorithms on inorganic crystal structures. \emph{Inorg. Chem.}
  \textbf{2021}, \emph{60}, 1590--1603\relax
\mciteBstWouldAddEndPuncttrue
\mciteSetBstMidEndSepPunct{\mcitedefaultmidpunct}
{\mcitedefaultendpunct}{\mcitedefaultseppunct}\relax
\EndOfBibitem
\bibitem[Kingma and Ba(2014)Kingma, and Ba]{kingma2014adam}
Kingma,~D.~P.; Ba,~J. Adam: A method for stochastic optimization. \emph{arXiv
  preprint arXiv:1412.6980} \textbf{2014}, \relax
\mciteBstWouldAddEndPunctfalse
\mciteSetBstMidEndSepPunct{\mcitedefaultmidpunct}
{}{\mcitedefaultseppunct}\relax
\EndOfBibitem
\bibitem[Bl{\"o}chl(1994)]{blochl1994projector}
Bl{\"o}chl,~P.~E. Projector augmented-wave method. \emph{Phys. Rev. B}
  \textbf{1994}, \emph{50}, 17953\relax
\mciteBstWouldAddEndPuncttrue
\mciteSetBstMidEndSepPunct{\mcitedefaultmidpunct}
{\mcitedefaultendpunct}{\mcitedefaultseppunct}\relax
\EndOfBibitem
\bibitem[Fletcher and Reeves(1964)Fletcher, and Reeves]{fletcher1964function}
Fletcher,~R.; Reeves,~C.~M. Function minimization by conjugate gradients.
  \emph{Comput. J.} \textbf{1964}, \emph{7}, 149--154\relax
\mciteBstWouldAddEndPuncttrue
\mciteSetBstMidEndSepPunct{\mcitedefaultmidpunct}
{\mcitedefaultendpunct}{\mcitedefaultseppunct}\relax
\EndOfBibitem
\bibitem[Larsen \latin{et~al.}(2017)Larsen, Mortensen, Blomqvist, Castelli,
  Christensen, Du{\l}ak, Friis, Groves, Hammer, Hargus, \latin{et~al.}
  others]{larsen2017atomic}
Larsen,~A.~H.; Mortensen,~J.~J.; Blomqvist,~J.; Castelli,~I.~E.;
  Christensen,~R.; Du{\l}ak,~M.; Friis,~J.; Groves,~M.~N.; Hammer,~B.;
  Hargus,~C.; others The atomic simulation environment—a Python library for
  working with atoms. \emph{J. Condens. Matter Phys.} \textbf{2017}, \emph{29},
  273002\relax
\mciteBstWouldAddEndPuncttrue
\mciteSetBstMidEndSepPunct{\mcitedefaultmidpunct}
{\mcitedefaultendpunct}{\mcitedefaultseppunct}\relax
\EndOfBibitem
\bibitem[Nocedal(1980)]{nocedal1980updating}
Nocedal,~J. Updating quasi-Newton matrices with limited storage. \emph{Math.
  Comput.} \textbf{1980}, \emph{35}, 773--782\relax
\mciteBstWouldAddEndPuncttrue
\mciteSetBstMidEndSepPunct{\mcitedefaultmidpunct}
{\mcitedefaultendpunct}{\mcitedefaultseppunct}\relax
\EndOfBibitem
\bibitem[Plimpton(1995)]{plimpton1995fast}
Plimpton,~S. Fast parallel algorithms for short-range molecular dynamics.
  \emph{J. Comput. Phys.} \textbf{1995}, \emph{117}, 1--19\relax
\mciteBstWouldAddEndPuncttrue
\mciteSetBstMidEndSepPunct{\mcitedefaultmidpunct}
{\mcitedefaultendpunct}{\mcitedefaultseppunct}\relax
\EndOfBibitem
\bibitem[Thompson \latin{et~al.}(2022)Thompson, Aktulga, Berger, Bolintineanu,
  Brown, Crozier, in't Veld, Kohlmeyer, Moore, Nguyen, \latin{et~al.}
  others]{thompson2022lammps}
Thompson,~A.~P.; Aktulga,~H.~M.; Berger,~R.; Bolintineanu,~D.~S.; Brown,~W.~M.;
  Crozier,~P.~S.; in't Veld,~P.~J.; Kohlmeyer,~A.; Moore,~S.~G.; Nguyen,~T.~D.;
  others LAMMPS-a flexible simulation tool for particle-based materials
  modeling at the atomic, meso, and continuum scales. \emph{Comput. Phys.
  Commun.} \textbf{2022}, \emph{271}, 108171\relax
\mciteBstWouldAddEndPuncttrue
\mciteSetBstMidEndSepPunct{\mcitedefaultmidpunct}
{\mcitedefaultendpunct}{\mcitedefaultseppunct}\relax
\EndOfBibitem
\bibitem[Qi \latin{et~al.}(2024)Qi, Ko, Wood, Pham, and Ong]{qi2024robust}
Qi,~J.; Ko,~T.~W.; Wood,~B.~C.; Pham,~T.~A.; Ong,~S.~P. Robust training of
  machine learning interatomic potentials with dimensionality reduction and
  stratified sampling. \emph{npj Comput. Mater.} \textbf{2024}, \emph{10},
  43\relax
\mciteBstWouldAddEndPuncttrue
\mciteSetBstMidEndSepPunct{\mcitedefaultmidpunct}
{\mcitedefaultendpunct}{\mcitedefaultseppunct}\relax
\EndOfBibitem
\bibitem[Zhang \latin{et~al.}(1996)Zhang, Ramakrishnan, and
  Livny]{zhang1996birch}
Zhang,~T.; Ramakrishnan,~R.; Livny,~M. BIRCH: an efficient data clustering
  method for very large databases. \emph{SIGMOD Rec.} \textbf{1996}, \emph{25},
  103--114\relax
\mciteBstWouldAddEndPuncttrue
\mciteSetBstMidEndSepPunct{\mcitedefaultmidpunct}
{\mcitedefaultendpunct}{\mcitedefaultseppunct}\relax
\EndOfBibitem
\bibitem[Poli{\v{c}}ar \latin{et~al.}(2019)Poli{\v{c}}ar, Stra{\v{z}}ar, and
  Zupan]{polivcar2019opentsne}
Poli{\v{c}}ar,~P.~G.; Stra{\v{z}}ar,~M.; Zupan,~B. openTSNE: a modular Python
  library for t-SNE dimensionality reduction and embedding. \emph{BioRxiv}
  \textbf{2019}, 731877\relax
\mciteBstWouldAddEndPuncttrue
\mciteSetBstMidEndSepPunct{\mcitedefaultmidpunct}
{\mcitedefaultendpunct}{\mcitedefaultseppunct}\relax
\EndOfBibitem
\bibitem[Linderman \latin{et~al.}(2019)Linderman, Rachh, Hoskins,
  Steinerberger, and Kluger]{linderman2019fast}
Linderman,~G.~C.; Rachh,~M.; Hoskins,~J.~G.; Steinerberger,~S.; Kluger,~Y. Fast
  interpolation-based t-SNE for improved visualization of single-cell RNA-seq
  data. \emph{Nat. Methods} \textbf{2019}, \emph{16}, 243--245\relax
\mciteBstWouldAddEndPuncttrue
\mciteSetBstMidEndSepPunct{\mcitedefaultmidpunct}
{\mcitedefaultendpunct}{\mcitedefaultseppunct}\relax
\EndOfBibitem
\end{mcitethebibliography}
\end{document}


\newpage
\section{Silicon}

\begin{figure}[H]
    \centering
    \includegraphics[width=0.5\textwidth]{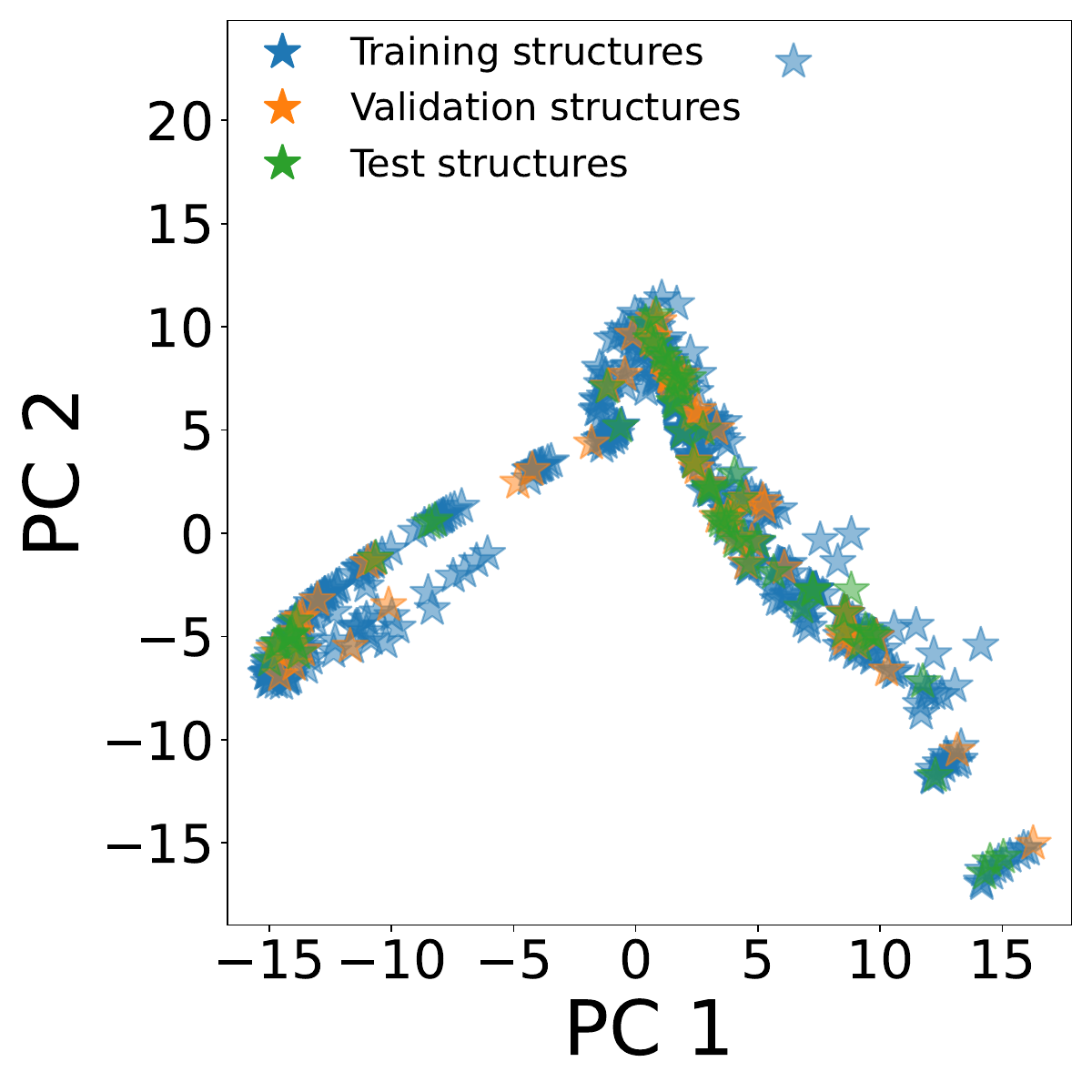}
    \caption{\textbf{Plot of first two principal components of the M3GNet-encoded structure features for training, validation and test sets of silicon.}}
    \label{fig:Si_coverage}
\end{figure}

\begin{table}
    \centering
\caption{\textbf{DIRECT sampling hyperparameters for silicon.} These include the threshold, $n$, and $k$ are used to construct 1fi and mfi M3GNet potentials.}
    \begin{tabular}{c c c c}
    property & threshold & $n$ & $k$  \\
    \hline\hline
 1\%-mfi SCAN & 0.05 & 6 & 1 \\
 5\%-mfi SCAN & 0.05 & 30 & 1 \\
 10\%-mfi SCAN  & 0.05 &  60 & 1  \\
 20\%-mfi SCAN & 0.05 & 63 & 2 \\
 50\%-mfi SCAN & 0.05 & 205 & 2 \\
    \hline
    \end{tabular}
    \label{tab:DIRECT_settings}
\end{table} 

\begin{figure}[H]
    \centering
    \includegraphics[width=1.0\textwidth]{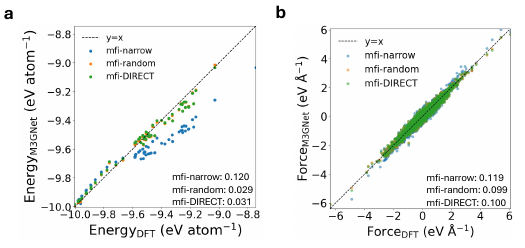}
    \caption{\textbf{Parity plots of M3GNet vs DFT energies and forces for silicon.} Parity plots of \textbf{a} energies and \textbf{b} forces predicted by 10\%-mfi-narrow, 10\%-mfi-random and 10\%-mfi-DIRECT M3GNet GLPs. The numbers indicate the mean absolute error (MAE) in eV atom$^{-1}$ and eV \si{\angstrom}$^{-1}$ for energies and forces, respectively.}
    \label{fig:Si_error}
\end{figure}

\begin{figure}
    \centering
    \includegraphics[width=1.0\textwidth]{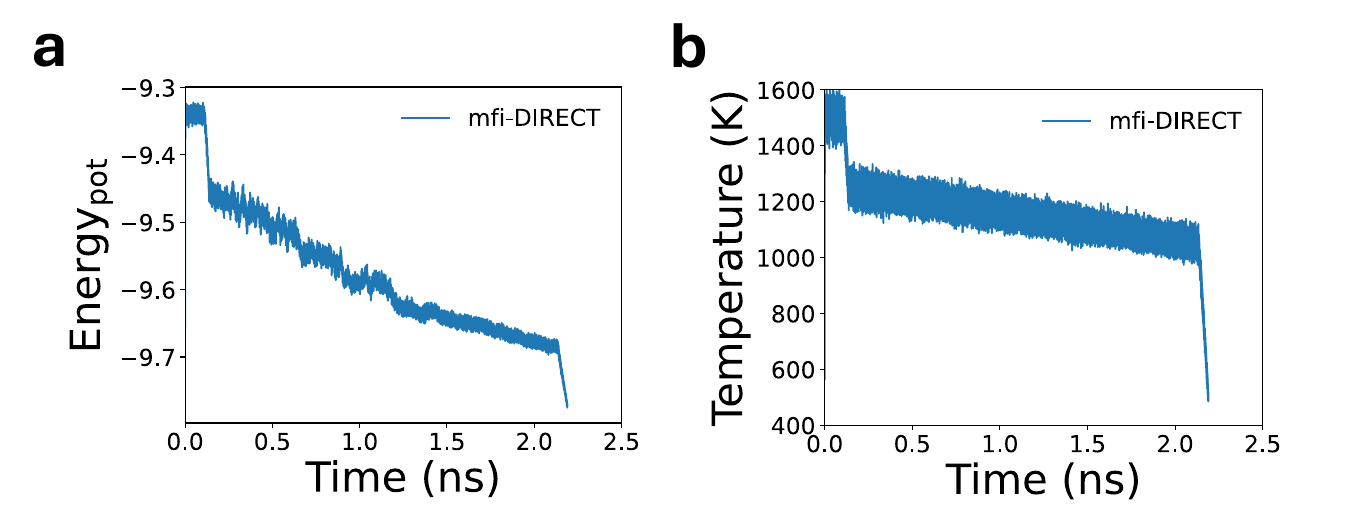}
    \caption{\label{fig:Si_melt_and_quench}\textbf{Temperature and potential energy of silicon during melt-and-quench simulation.} The simulated temperature and the corresponding potential energy of silicon were obtained by starting from initially relaxed structures, then equilibrating at 1500K and quenching down to 500K in three steps. A fast quenching rate was used for the temperature ranges from 1500K to 1250K and from 1050K to 500K, while a slow quenching rate was applied to the temperature range from 1250K to 1050K.}
    
\end{figure}

\section{Water}
\begin{figure}[H]
    \centering
    \includegraphics[width=1.0\textwidth]{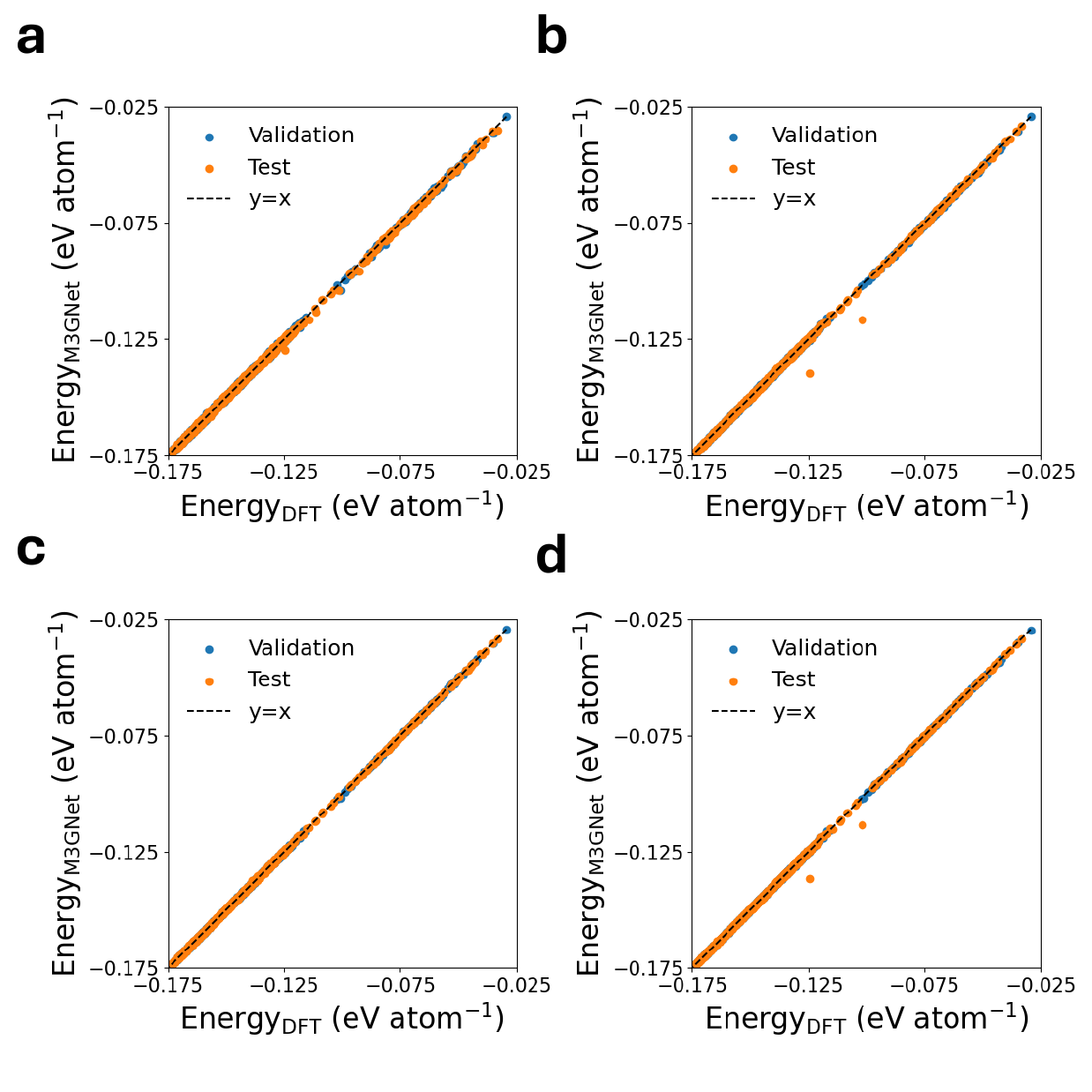}
    \caption{\textbf{Parity plots of M3GNet vs DFT energies for water.} The normalized energy of structures from the training and validation set is computed using (a) 10$\%$-1fi SCAN, (b) 50$\%$-1fi SCAN, (c) 80$\%$-1fi SCAN and (d) 10$\%$-mfi SCAN models and DFT. The energy values are shifted by adding 156 eV atom$^{-1}$ into both M3GNet and DFT predictions for better visualization.}
    \label{fig:H2O_energy}
\end{figure}

\begin{figure}[H]
    \centering
    \includegraphics[width=1.0\textwidth]{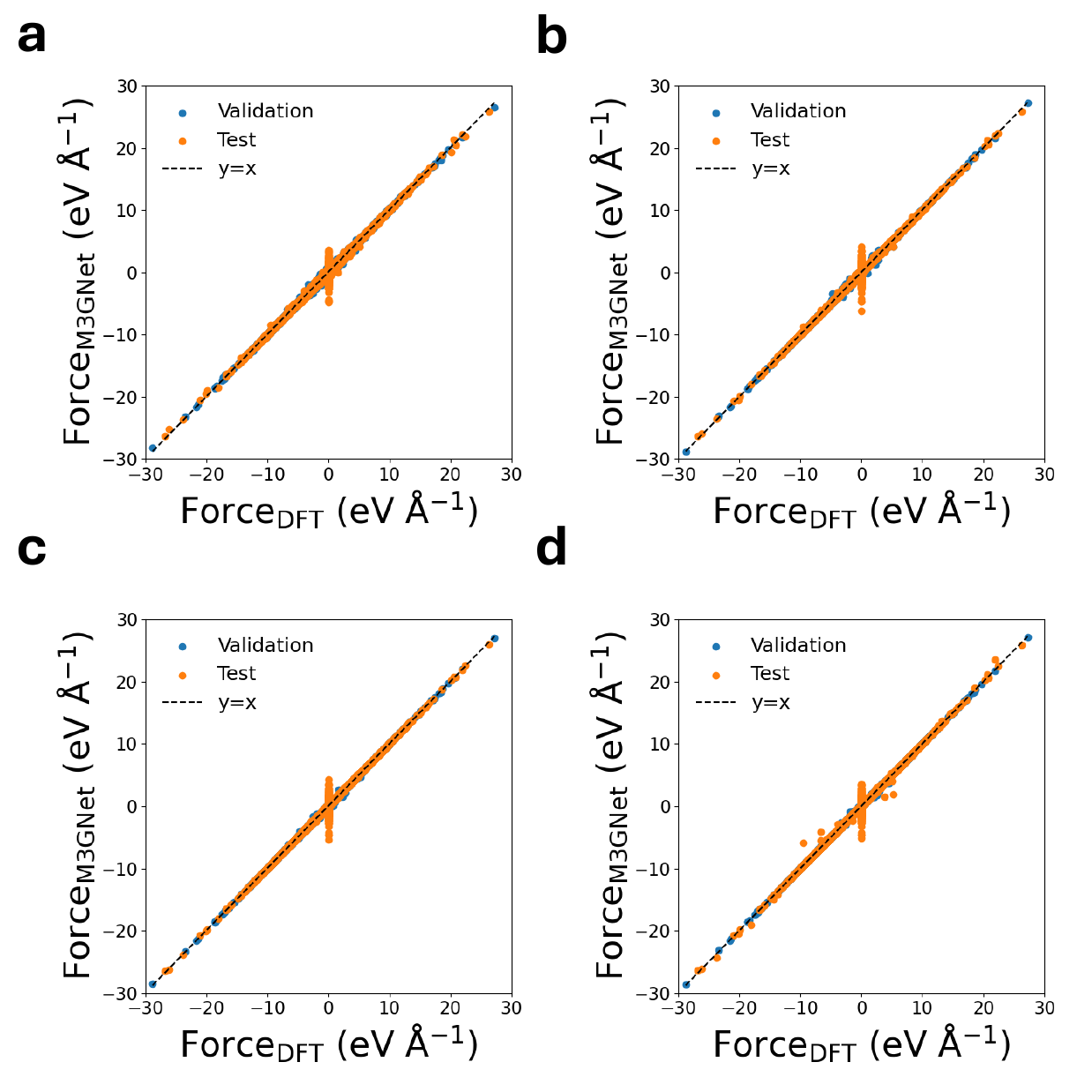}
    \caption{\textbf{Parity plots of M3GNet vs DFT forces for water.} The atomic force component of structures from the training and validation set is computed using (a) 10$\%$-1fi SCAN, (b) 50$\%$-1fi SCAN, (c) 80$\%$-1fi SCAN and (d) 10$\%$-mfi SCAN models and DFT. }
    \label{fig:H2O_force}
\end{figure}

\begin{figure}[H]
    \centering
    \includegraphics[width=0.5\textwidth]{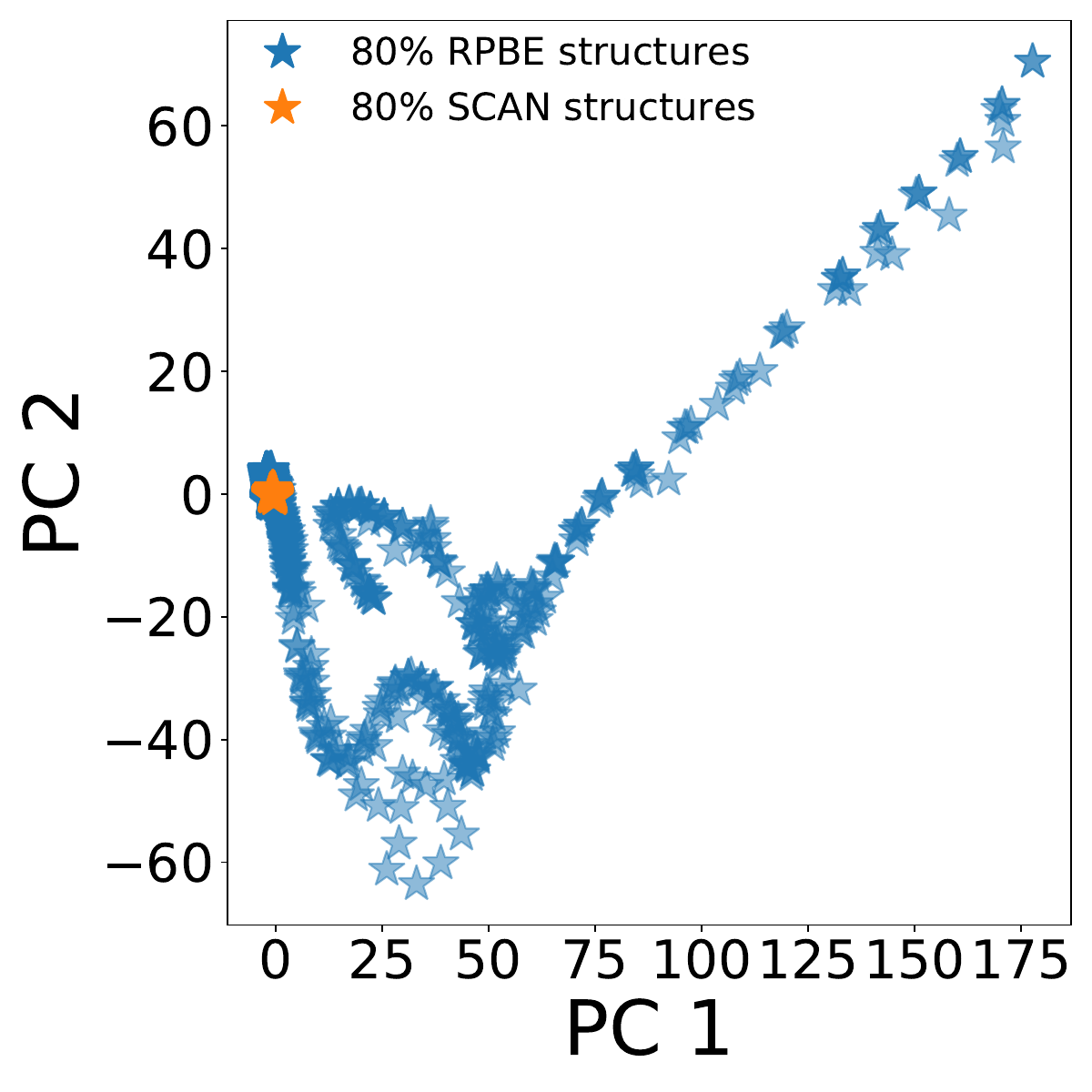}
    \caption{\textbf{Plot of first two principal components of the M3GNet-encoded structure features for water.} The configuration space covered by the 80\% of SCAN structures (blue star) used for training either 1fi or mfi M3GNet models are within the RPBE structures (orange star).}
    \label{fig:H2O_coverage}
\end{figure}

\begin{figure}[H]
    \centering
    \includegraphics[width=1.0\textwidth]{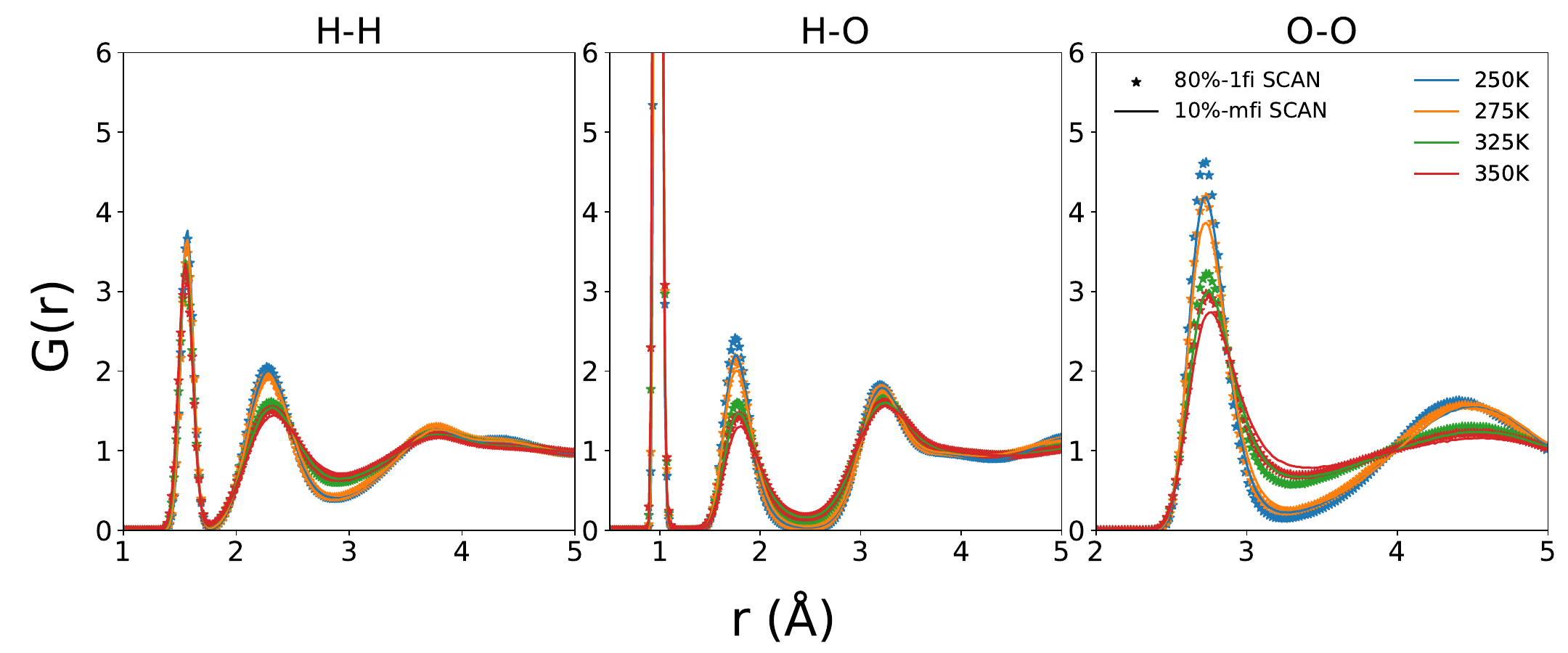}
    \caption{\textbf{Radial distribution functions of liquid water at different temperatures.} The radial distribution functions of liquid water at 250, 275, 325 and 350K were obtained from \textit{NPT} simulations for 0.5 ns under 1 atm pressure.}
    \label{fig:H2O_RDF_si}
\end{figure}
\begin{figure}[H]
    \centering
    \includegraphics[width=0.75\textwidth]{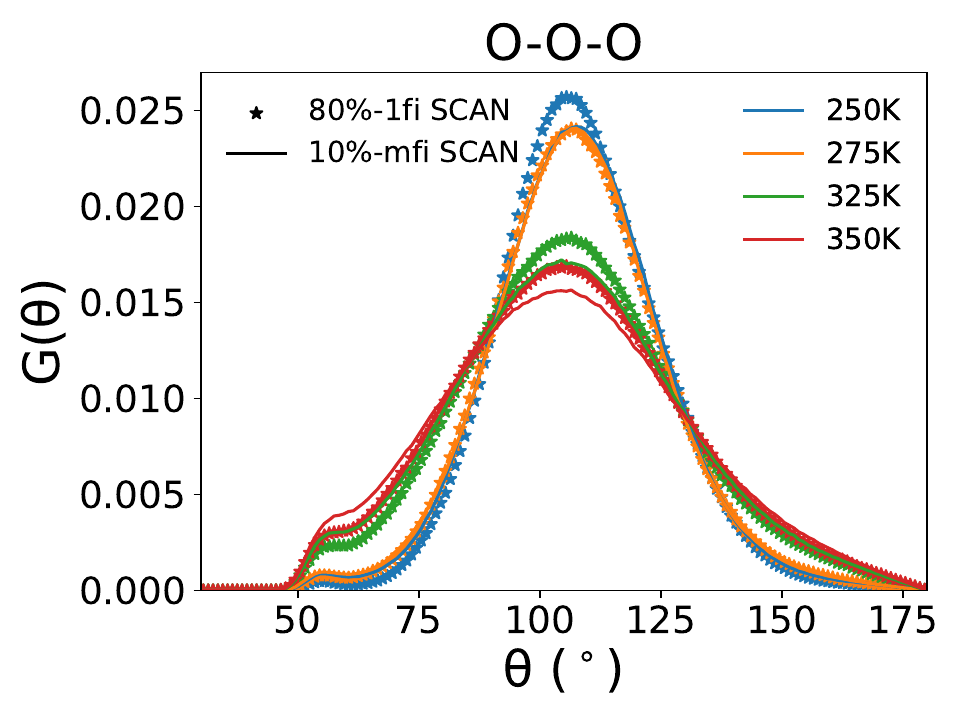}
    \caption{\textbf{Angular distribution functions of liquid water at different temperatures.} The angular distribution functions of liquid water at 250, 275, 325 and 350K were obtained from \textit{NPT} simulations for 0.5 ns under 1 atm pressure.}
    \label{fig:H2O_ADF_si}
\end{figure}